\documentclass[twocolumn]{aastex63}
\usepackage[utf8]{inputenc}
\usepackage{times}
\usepackage{amsmath,amsthm,amssymb}
\usepackage{graphics, graphicx}
\usepackage{multirow}

\shorttitle{SALT Spectropolarimetry of FSRQs}
\shortauthors{S. Podjed et al.}

\begin{document}
\title{Optical Spectropolarimetric Variability Properties in Blazars PKS 0637--75 and PKS 1510--089 }

\author[0000-0002-0504-565X]{Stephanie A. Podjed}
\affiliation{Department of Physics and Astronomy, Dartmouth College, 6127 Wilder Laboratory, Hanover, NH 03755-3528}

\author[0000-0003-1468-9526]{Ryan C. Hickox}
\affiliation{Department of Physics and Astronomy, Dartmouth College, 6127 Wilder Laboratory, Hanover, NH 03755-3528}

\author[0000-0003-4042-2438]{Jedidah C. Isler}
\affiliation{The SeRCH Foundation, Inc, PO Box 442335, Fort Washington, MD 20749-2335}

\author[0000-0002-8434-5692]{Markus Böttcher}
\affiliation{Centre for Space Research, North-West University, Potchefstroom 2520, South Africa}

\author[0000-0002-1769-5617]{Hester M. Schutte}
\affiliation{Centre for Space Research, North-West University, Potchefstroom 2520, South Africa}

\correspondingauthor{Stephanie Podjed}
\email{stephanie.a.podjed.gr@dartmouth.edu}

\received{2024 February 8}
\revised{2024 April 16}
\accepted{2024 April 18}
\published{2024 June 20}
\submitjournal{The Astrophysical Journal}

\begin{abstract}

\noindent Spectropolarimetry is a powerful tool to investigate the central regions of active galactic nuclei (AGNs), as polarization signatures are key to probing magnetic field structure, evolution, and the physics of particle acceleration in jets. Optical linear polarization of blazars is typically greater than a few percent, indicating the emission is dominated by non-thermal synchrotron radiation, while polarization less than a few percent is common for other type 1 AGNs. We present a spectropolarimetric study of PKS 0637--75 and PKS 1510--089 to determine how the head-on orientation of a jet and dominant emission processes influence polarimetric variations in the broad lines and continuum. Observations were obtained bi-weekly from the Robert Stobie Spectrograph on the Southern African Large Telescope. Variability in the continuum polarization is detected for both PKS 0637--75 and PKS 1510--089, with a total average level of 2.5\% $\pm$ 0.1\% and 7.5\% $\pm$ 0.1\%, respectively. There is no clear polarization in the broad Balmer emission lines and weak polarization in Mg~II as the average level across all observations is 0.2\% $\pm$ 0.1\% for H$\beta$, 0.2\% $\pm$ 0.3\% for H$\gamma$, and 0.6\% $\pm$ 0.2\% for Mg~II. We find that polarization measurements confirm the conclusions drawn from spectral energy distribution modeling of the disk-jet contributions to the emission, as optical polarization and time variability for PKS 0637--75 is shown to be dominated by accretion disk emission while that of PKS 1510--089 is due to both disk and jet emission, with greater jet contribution during flaring states.

\end{abstract}


\section{Introduction}

Blazars are a class of radio-loud (jet-dominated) active galactic nuclei (AGNs) whose relativistic jet is oriented at small angles with respect to our line of sight \citep{ant1993,urry,padovani17}. The bulk outflow of these charged particles, in addition to the small viewing angles, induces Doppler beaming along the direction of the outflow which makes the jet appear brighter and the variability timescale shorter in the observed frame. Blazars can be divided into two subclasses based on optical spectral features: Flat Spectrum Radio Quasars (FSRQs) and BL Lacertae objects (BL Lacs). The spectrum of an FSRQ generally has prominent broad and narrow emission lines, whereas that of a BL Lac object has absent or weak (EW $<$ 5 Å) emission lines \citep{urry}.

Blazar broadband emission experiences rapid variability and is generally jet-dominated. During quiescent periods the contribution of thermal emission from an optically-thick accretion disk can be detected, making blazars an ideal laboratory to study connections between the nonthermal jet and thermal disk components. Radio through optical, and more recently through X-ray with the Imaging
X-ray Polarimetry Explorer \citep{ixpe, ixpe_blazar3, ixpe_blazar5, ixpe_blazar1, ixpe_blazar2, ixpe_blazar4, ixpe_blazar6}, emission of blazars is also characterized by a high degree of polarization, with polarization percentages ranging from a few to tens of percent, while their non-beamed AGN counterparts usually show polarization levels of less than a few percent. The optical emission of blazars is often dominated by synchrotron radiation from the relativistic jet and is well-known to be polarized, with a theoretical maximum synchrotron polarization percentage attainable ranging from 69\% to 75\% \citep{ryli}. The polarization percentage (\textit{P}) and electric vector position angle are both often highly variable, even on intranight timescales \citep{darcangelo, zhang14}.

Polarimetry as a complement to spectroscopy is a useful tool to investigate the central regions of AGNs that are unresolved by direct observations since polarization is sensitive to the geometry and magnetic fields of the scattering region. Optical spectropolarimetric studies of type 2 AGNs  \citep{antmill85, millgoodwill} provided the basis for the phenomenological geometric unification scheme of active galactic nuclei \citep{ant1993, urry,netzer15}. Though the optical light coming from the central engine and broad-line region (BLR) in these objects is thought to be geometrically blocked by an obscuring torus, it can escape perpendicularly along the unobstructed direction by scattering off of dust or free electrons located above or below the central region \citep{ant1983,lira20}. This scattering induces linear polarization in the emission, making the once-hidden broad emission lines observable. Similar studies of type 1 AGNs, objects where the BLR is directly observable, found that the optical polarization properties are not consistent with this polar scattering; instead, the PA is typically aligned with the radio axis/axis of symmetry, suggesting that the scattering material is equatorially located \citep{ant1983, smith2002, smith2004, smith2005}.

The mechanisms responsible for polarization in the continuum and broad lines of AGNs can be different, and include the intrinsic geometry of the emitting region and the geometry of the scattering region \citep{afanasiev14}, as polarization due to radiative transfer alone cannot account for any polarization levels in the BLR \citep{afanasiev19}. Additionally in blazars, the thermal components, i.e., the AD, BLR, and dusty torus, are expected to be unpolarized and in the optical spectrum will be characterized by a decrease in the degree of polarization in spectropolarimetric observations \citep{Bot17}.

For blazars, which generally have a face-on geometry and unobscured view of the accretion flow, spectropolarimetry is a diagnostic tool that can be used to better understand the interplay between the nonthermal jet and thermal disk components. These findings can be used to comment on the presence and location of scattering regions, how blazar spectra respond to accretion disk dominated emission or synchrotron dominated emission, and to better constrain emission mechanism models when included in spectral energy distribution (SED) studies \citep{Bot17, schutte}.

FSRQs are extremely well studied in time series observations across a range of wavelengths, but a relative paucity of spectropolarimetric observations of blazars exist \citep[e.g.,][] {Bot17, patino, schutte, aharonian+23}, with such studies becoming prominent only recently. Since we aim to increase the characterization of blazar spectropolarimetric properties, we focus on FSRQs as they generally show prominent quasar-like emission lines in contrast to BL Lacs. Particularly in this work, from our sample of seven Southern Hemisphere gamma-ray active and quiescent blazars, we present the optical linear polarization variability properties for the two objects displaying emission lines: PKS 0637--75 (\textit{z} = 0.653, J2000 RA = 06$^{\mathrm{h}}$35$^{\mathrm{m}}$46$^{\mathrm{s}}$.5079, Dec = -75$^{\mathrm{d}}$16$^{\mathrm{m}}$16$^{\mathrm{s}}$.814 as given in the NASA/IPAC Extragalactic Database, NED\footnote{\url{https://ned.ipac.caltech.edu/}}) and PKS 1510--089 (\textit{z} = 0.36, J2000 RA = 15$^{\mathrm{h}}$12$^{\mathrm{m}}$50$^{\mathrm{s}}$.53, Dec = -09$^{\mathrm{d}}$05$^{\mathrm{m}}$59$^{\mathrm{s}}$.82 as given in NED). We focus on the Mg~II $\lambda$2798 emission line, a typically strong low ionization line seen in the optical-UV band of AGNs, in PKS 0637--75 and on the broad H$\gamma$ and H$\beta$ emission lines in PKS 1510--089 .

In this paper, the observations and data analysis of PKS 0637--75 and PKS 1510--089 are described in Section \ref{obs}. Results of our polarization variability study are presented in Section \ref{res} and discussed in Section \ref{dis}. Our main conclusions are summarized in Section \ref{con}.
\section{Observations and Data Reduction} \label{obs}

\subsection{Optical Spectropolarimetry}

Optical spectropolarimetrc observations have been performed with the Robert Stobie Spectrograph \citep[RSS;][]{RSS, rss2} on the Southern African Large Telescope \citep[SALT;][]{salt}, a 10 m class telescope located at the South African Astronomical Observatory near Sutherland, South Africa. The effective area of the telescope is constantly changing as a part of the SALT design, so accurate absolute flux calibration is not available. The data were collected using the RSS in the spectropolarimetry LINEAR mode with slit-width of 1$\arcsec$.25 between 2019 February and 2020 March for PKS 0637--75 and between 2019 March and 2021 August for PKS 1510--089.

\begin{figure*}[ht]
\gridline{\fig{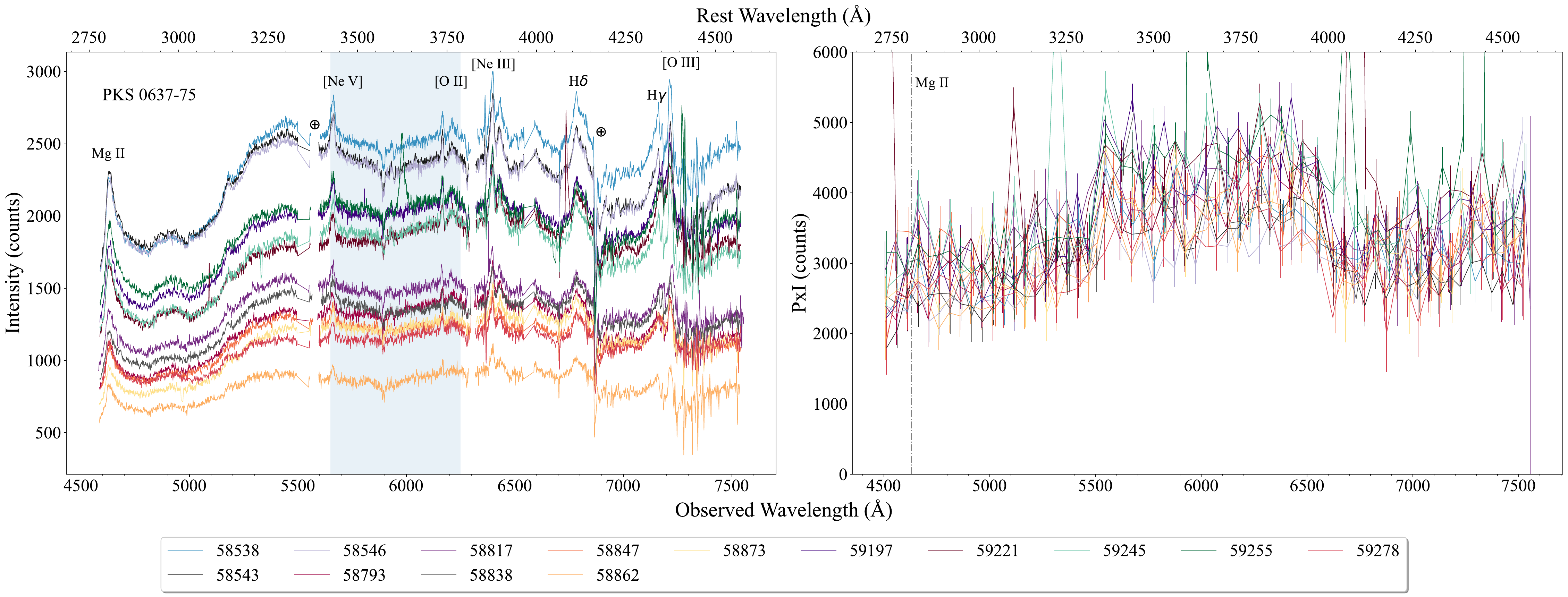}{\textwidth}{(a)}}
\gridline{\fig{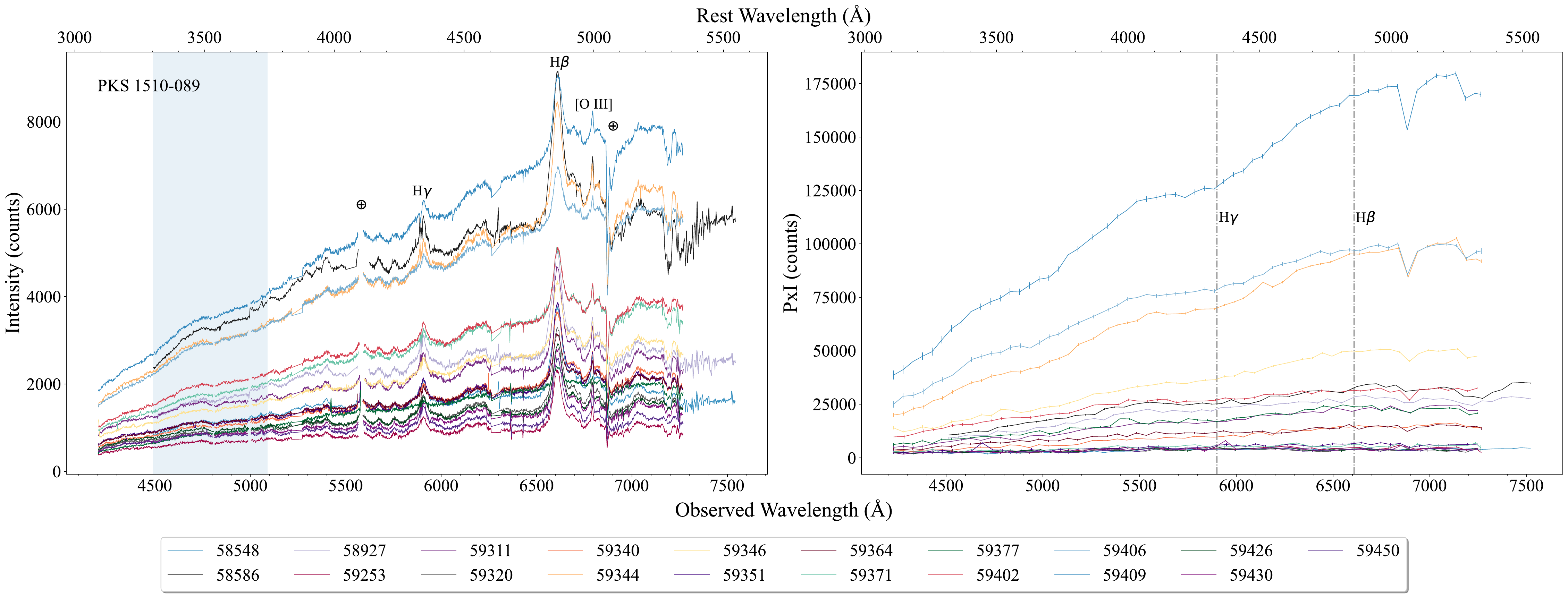}{\textwidth}{(b)}}

\caption{Optical SALT nonpolarized (left) and polarized (right) nonnormalized spectra of PKS 0637--75 (a) and PKS 1510--089   (b) obtained with the medium-resolution RSS. Identified emission and masked telluric ($\oplus$) lines are labeled. The continuum wavelength range used for analysis is highlighted. Gaps around 5500 and 6550 Å correspond to the gaps in the CCD as it is a mosaic of three individual chips. Vertical gray dotted--dashed lines in the polarized spectra indicate the location of the main emission lines studied here: Mg~II, H$\gamma$, and H$\beta$. Emission features in the nonpolarized emission are not detected in polarized light.}
\label{allspec}
\end{figure*}

\begin{figure}[ht]
\gridline{\fig{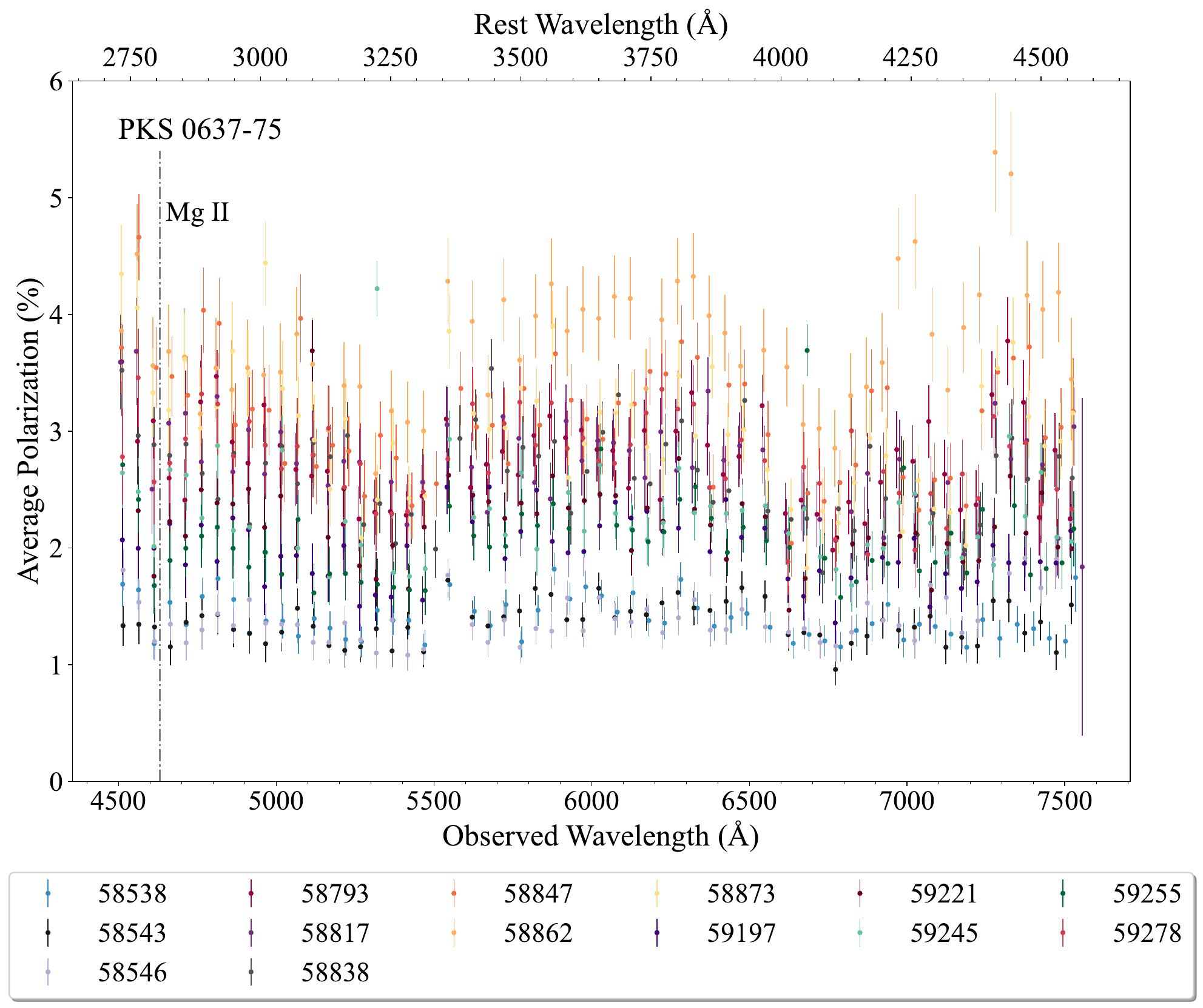}{0.5\textwidth}{(a)}}
\gridline{\fig{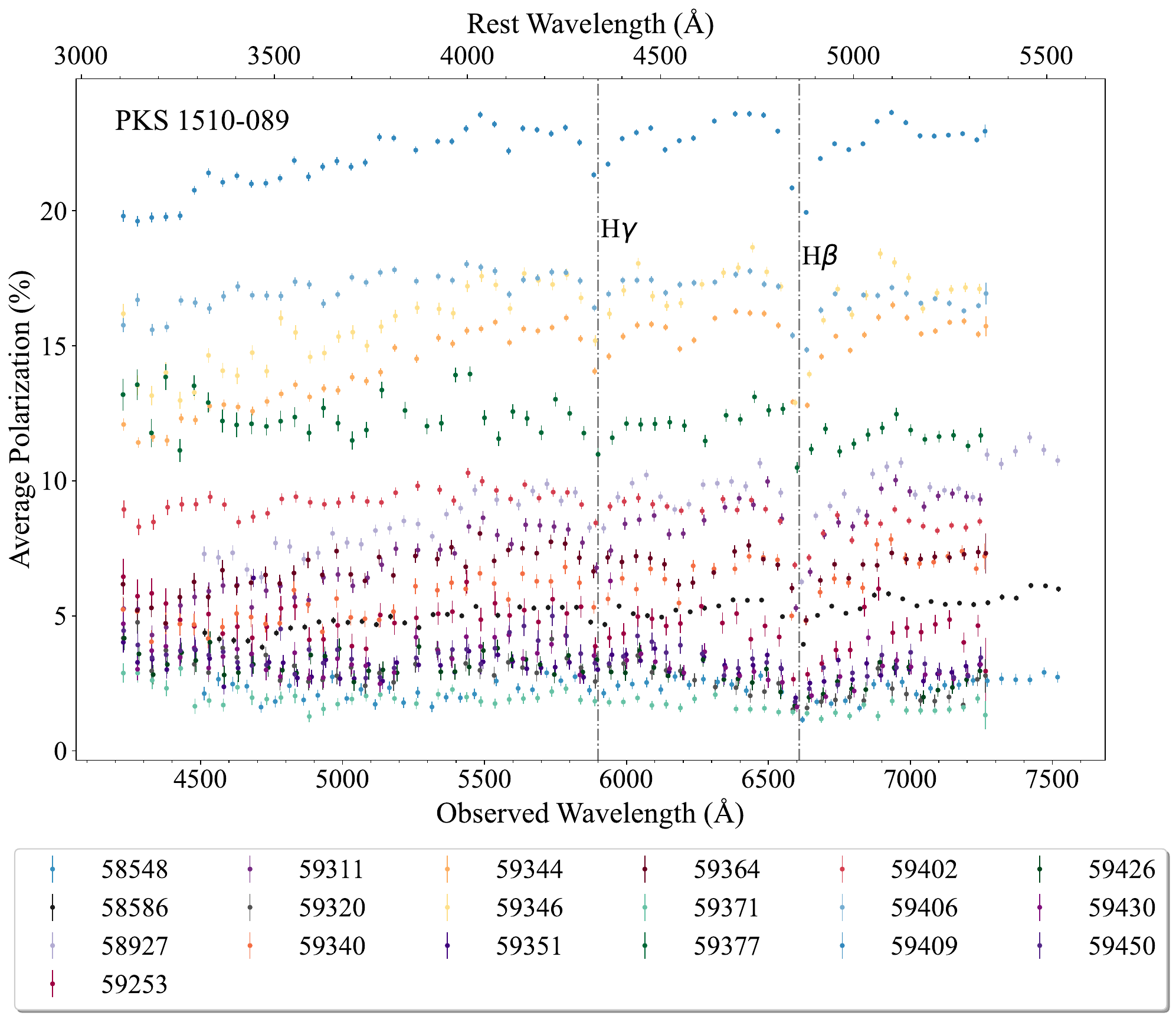}{0.5\textwidth}{(b)}}
\caption{Average polarization levels for PKS 0637--75 (a) and PKS 1510--089   (b); dotted--dashed gray lines mark the emission lines Mg II, H$\gamma$, and H$\beta$. Depolarization associated with the H$\gamma$ and H$\beta$ emission lines are noticeable in all observations for PKS 1510--809.}
\label{wapol_all}
\end{figure}

\begin{figure}[ht]
\gridline{\fig{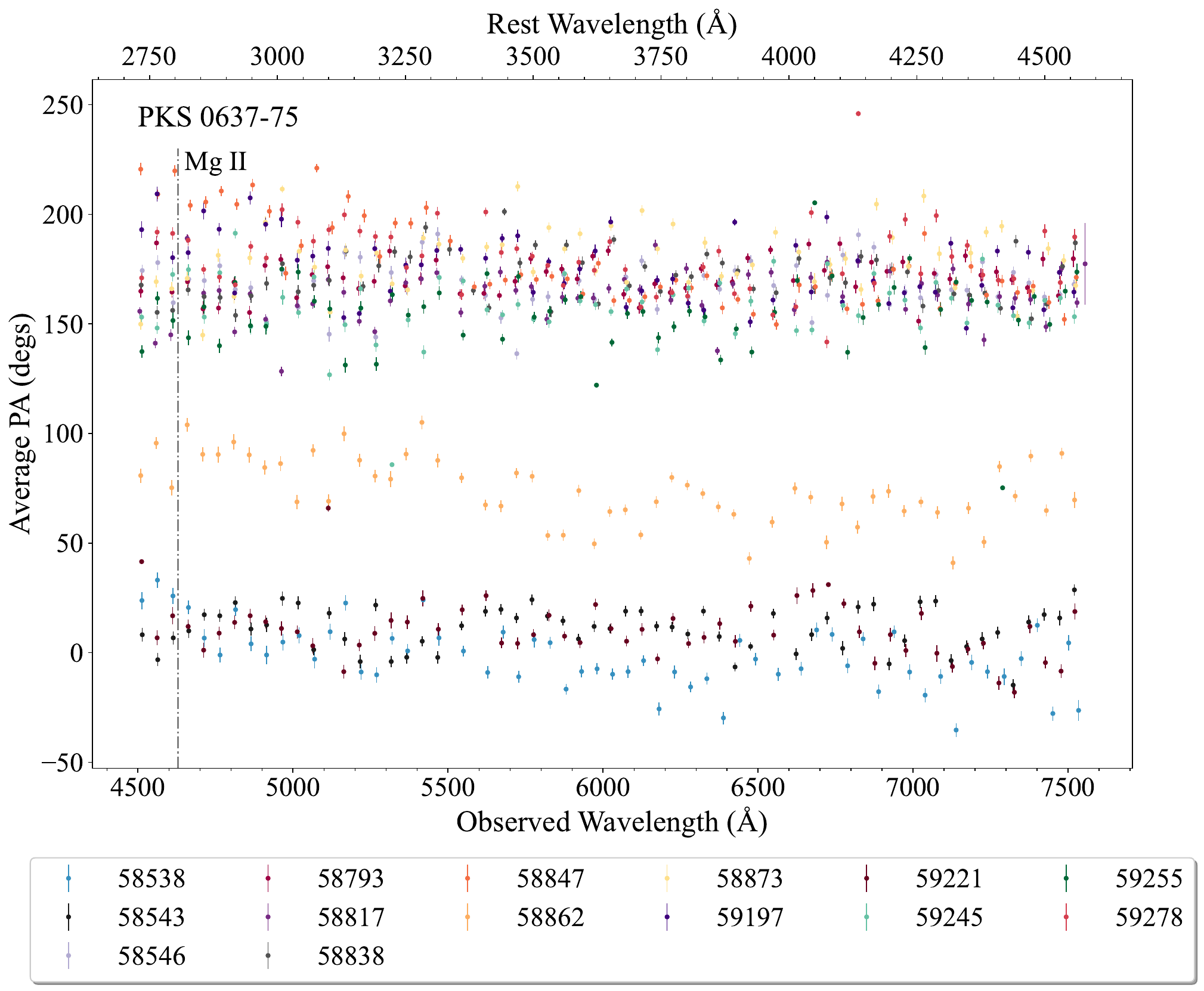}{0.5\textwidth}{(a)}}
\gridline{\fig{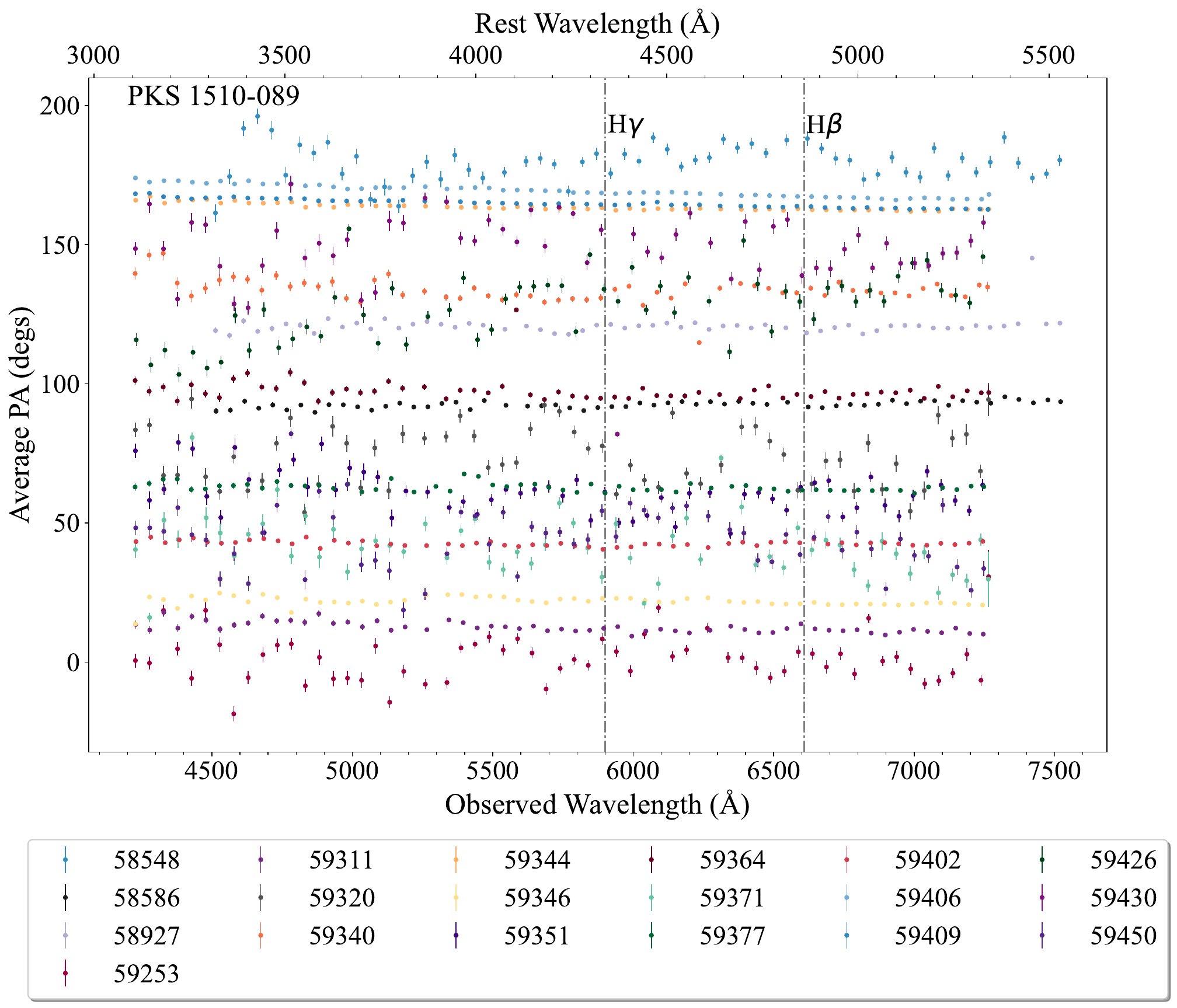}{0.5\textwidth}{(b)}}
\caption{Average polarization angle variability across wavelength and time of PKS 0637--75 (a) and PKS 1510--089   (b); dotted--dashed gray lines mark the emission lines Mg II, H$\gamma$, and H$\beta$. No PA swing across emission lines is seen for either object.}
\label{wapa}
\end{figure}

The RSS CCD detector consists of a mosaic of three CCDs with a total size of 6362 $\times$ 4102 pixels, with a single pixel size of $\sim$15 $\mu$m, corresponding to a spatial resolution of 0$\arcsec$.13 per pixel. We used the 2 $\times$ 4 binning, faint gain, slow readout mode. The mean gain of the mosaic is 1.7 ADU/electron and the read-out noise is typically around 2.48 electrons. We used the volume phase holographic (VPH) PG0900 grating with a grating angle of 15$^{\circ}$.88, which gives the observed wavelength coverage of 4500 Å -- 7500 Å. The average spectral resolving power is \textit{R} = 1300. Order blocking was done with the UV PC03850 filter. Each spectrum (flux, polarization degree, and polarization angle as a function of wavelength) was obtained through four total exposures per spectrum, one for each of the half-wave plate orientations (0$^{\circ}$, 45$^{\circ}$, 22.5$^{\circ}$, and 67.5$^{\circ}$) necessary to obtain linear polarization data. We began with an exposure time of 325 seconds that we reduced to 250 seconds for each subsequent observation after the first three. Each observing block consisted of flat-field images, slit acquisition images, the four science exposures, and an exposure of the calibration Xe lamp. See Figures \ref{allspec}, \ref{wapol_all}, and \ref{wapa} for the flux, average polarization, and average PA variability respectively.

\subsection{Optical and Gamma-Ray Photometry}
\begin{figure}[ht]
    \centering
    \includegraphics[width = \columnwidth]{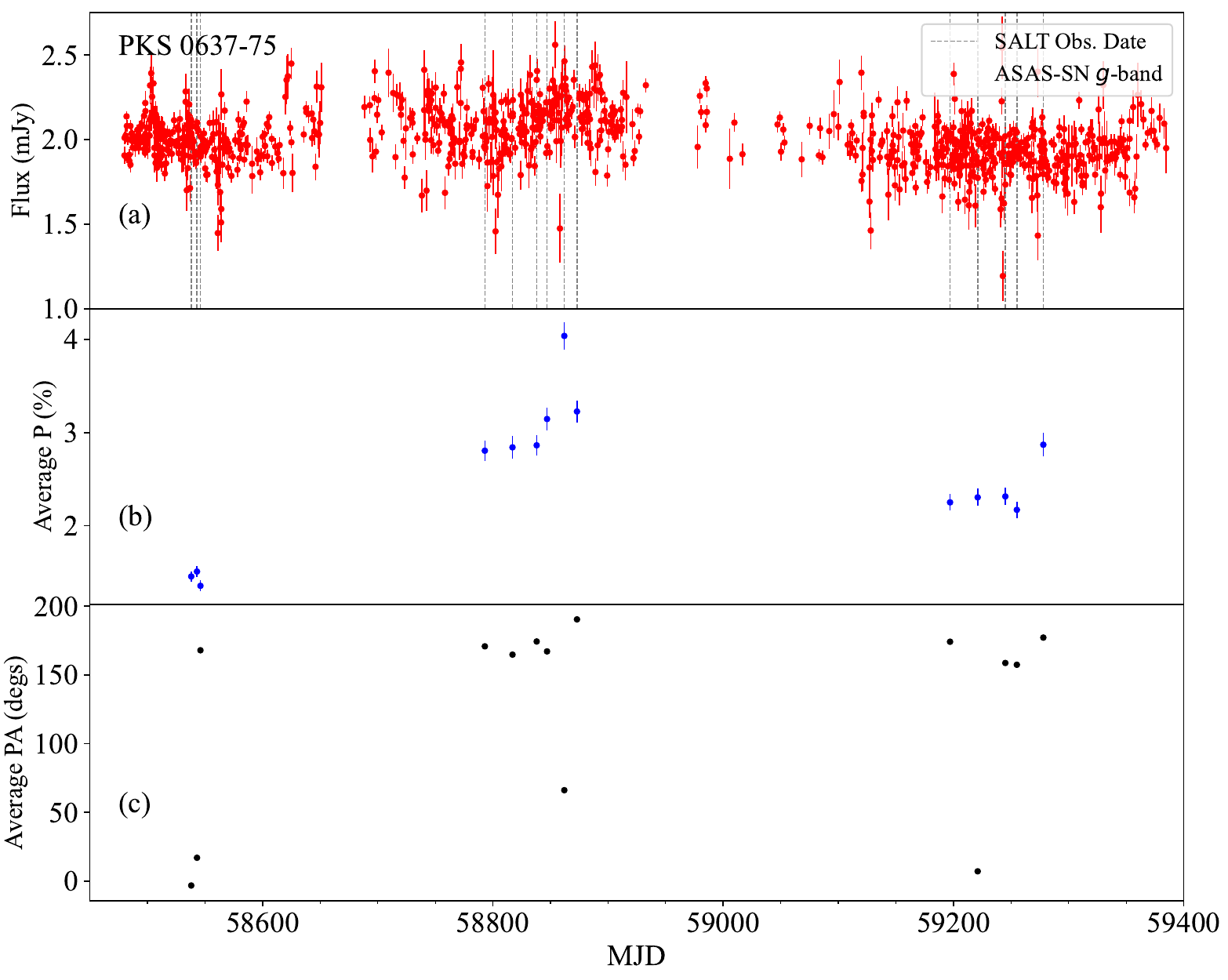}
    \caption{PKS 0637--75 ASAS-SN photometric \textit{g}-band flux corresponding to the period of our SALT spectropolarimetric study. Upper limits were removed for plotting purposes. Vertical gray dashed lines represent the MJD of specific nights of SALT observations. Panels (b) and (c) are the average continuum polarization and PA measurements respectively. The lack of strong optical flares (almost constant optical flux state) as well as low polarization levels during the epoch of observation may be indicating disk dominance of the emission and generally low to no disk activity in PKS 0637--75.}
    \label{0637asassn}
\end{figure}

\begin{figure*}[ht]
    \centering
    \includegraphics[width = \textwidth]{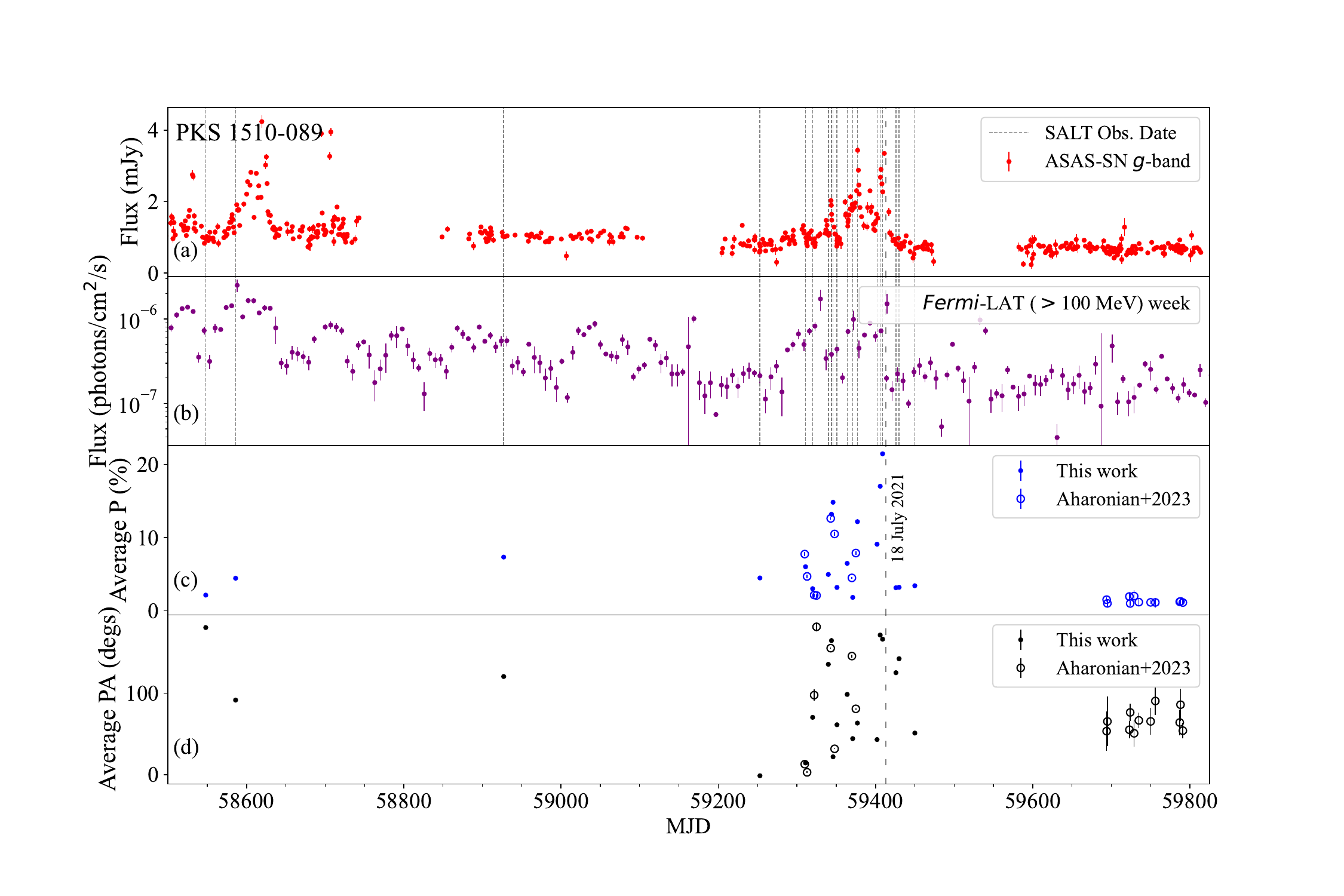}
    \caption{PKS 1510--089 ASAS-SN photometric \textit{g}-band flux (a) and \textit{Fermi} gamma-ray flux (b) corresponding to the period of obtained SALT spectropolarimetric data. Upper limits were removed for plotting purposes. Vertical gray dashed lines represent the MJD of specific nights of SALT observations. Panels (c) and (d) are the average continuum polarization and PA SALT measurements from this work and from \cite{aharonian+23}. The noticeable optical flare around MJD 59400 with high levels of polarization are indicating the jet dominance of the emission even when we don't have a  gamma-ray flare. Between 59260 to 59900 MJD, we can see that our SALT polarization and PA observations follow the trends found in \cite{aharonian+23} and provide more detail to the drop in polarization of PKS 1510--089 after 2021 July 18 (MJD 59413; vertical gray loosely dashed line) to an almost constant value of 2\%--3\%.}
    \label{1510_ahar_horiz}
\end{figure*}

We supplement our spectropolarimetric observations with \textit{g}-band photometry from the Ohio State All-Sky Automated Survey for Supernovae (ASAS-SN\footnote{\url{http://www.astronomy.ohio-state.edu/~assassin/index.shtml}}) project \citep{shappee, kochanek}, in order to monitor any continuum changes. Thermal continuum emission in the ultraviolet and optical is thought to originate within the AD \citep{perlman} and it is possible to have nonthermal continuum emission originating from synchrotron emission from the jet (see panel (a) of Figures \ref{0637asassn} and \ref{1510_ahar_horiz}). Additionally, we use the weekly binned 100--300,000 MeV \textit{Fermi} Large Area Telescope (\textit{Fermi}-LAT) light curve data for continually monitored sources\footnote{\url{https://fermi.gsfc.nasa.gov/ssc/data/access/lat/msl_lc/}} to monitor the jet behavior. After converting from mission elapsed time to MJD, we trimmed the data to cover a date range similar to that of our SALT data for PKS 1510--089   (panel (b) of Figure~\ref{1510_ahar_horiz}).

\subsection{Data Reduction}

Data analysis was performed using the graphical user interface (GUI) of the \texttt{polsalt}\footnote{\url{https://github.com/saltastro/polsalt}} reduction pipeline \citep{polsalt} for Python 2.7. \texttt{reducepoldataGUI.py} is a step-wise execution for reducing RSS spectropolarimetric data. It begins with raw image reductions, where basic CCD reduction techniques are undertaken including overscan subtraction, gain and cross talk correction, and amplifier mosaicing. Once this is finished, wavelength calibration is done for the two beams of the beamsplitter using the \texttt{IRAF specidentify} interface to identify the emission lines from the calibration arc images; cosmic ray rejection happens here as well. Next, in each image the beams are corrected for beamsplitter distortion and tilt, and the sky and target spectrum is extracted versus wavelength. Raw and final Stokes calculations are completed for the polarimetry reduction. In the raw Stokes calculation, wave-plate position pairs are identified and together result in linear polarization signal swapping between the \textit{O} and \textit{E} beams. The ``raw Stokes'' files contain unnormalized \textit{I} and \textit{S} plane data, with the degree of polarization being \textit{S/I}. In the final Stokes calculation, the full polarization pattern is evaluated to determine \textit{Q} and \textit{U}. Finally, polarimetric zero point, wave-plate efficiency, and axis calibrations are applied to give final Stokes parameters. Once all of these reduction steps are accomplished, the GUI provides an interactive results visualization window showing plots of the intensity, linear polarization percentage, equatorial position angle, Stokes \textit{Q} or Stokes \textit{U} behavior. 

In analyzing the SALT spectropolarimetric data we take a weighted average of the polarization (\textit{P}) PA data in 50 Å wide bins to investigate any polarization variability associated with each object in its continuum and broad emission lines. Error associated with these is the standard error on the weighted mean; see equations directly below, where \textit{x} is the parameter of which we find the weighted average:

\begin{equation}
    \Bar{x} = \frac{\Sigma \: x_i / \sigma_{x_i}^2}{\Sigma \:1/\sigma_{x_i}^2}\: , \: \: \:\sigma_{\Bar{x}} = \sqrt{\frac{1}{\Sigma \: 1/\sigma_{x_i}^2}}
\end{equation}

\begin{figure}[ht]
\gridline{\fig{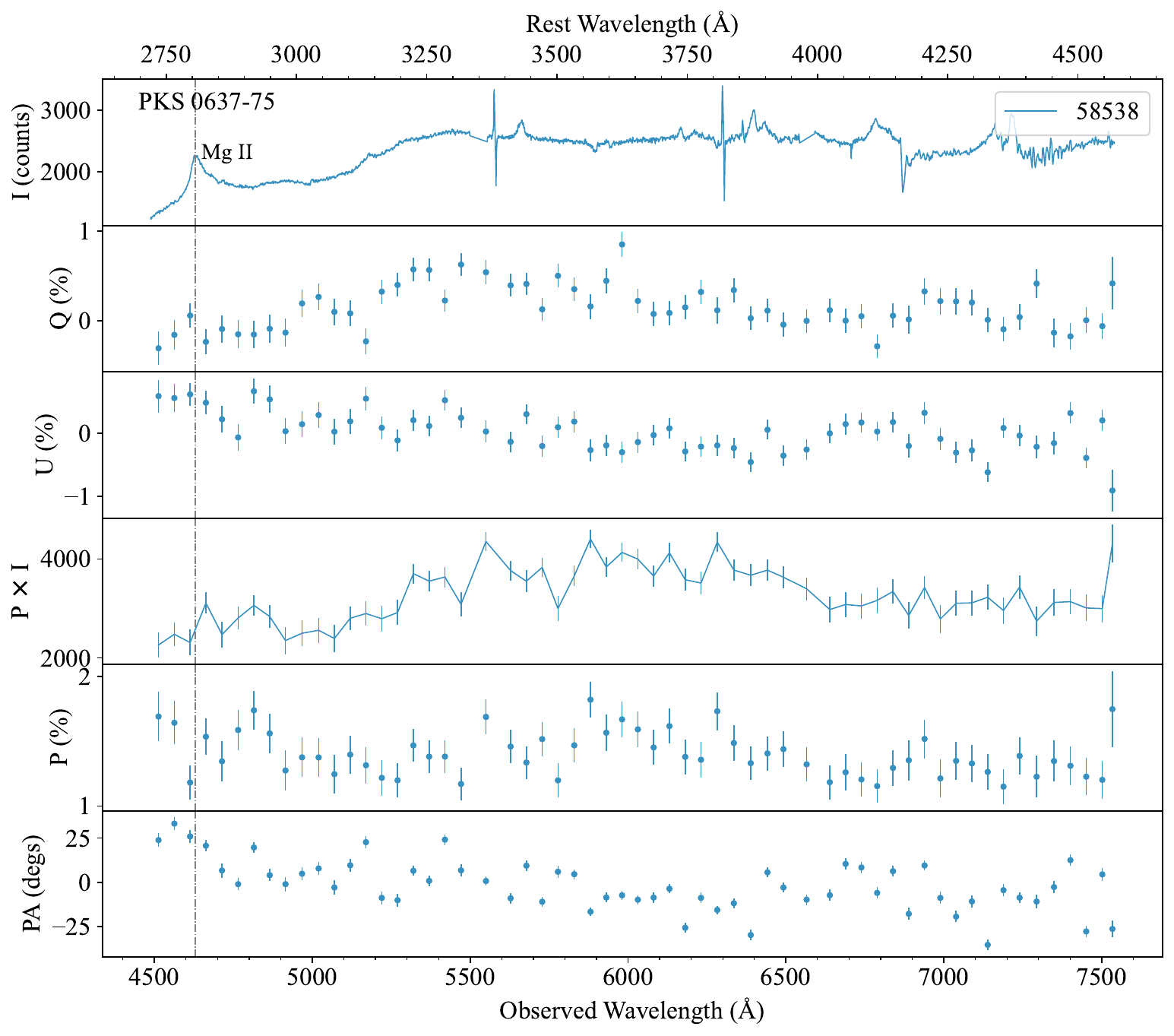}{\columnwidth}{(a)}}
\gridline{\fig{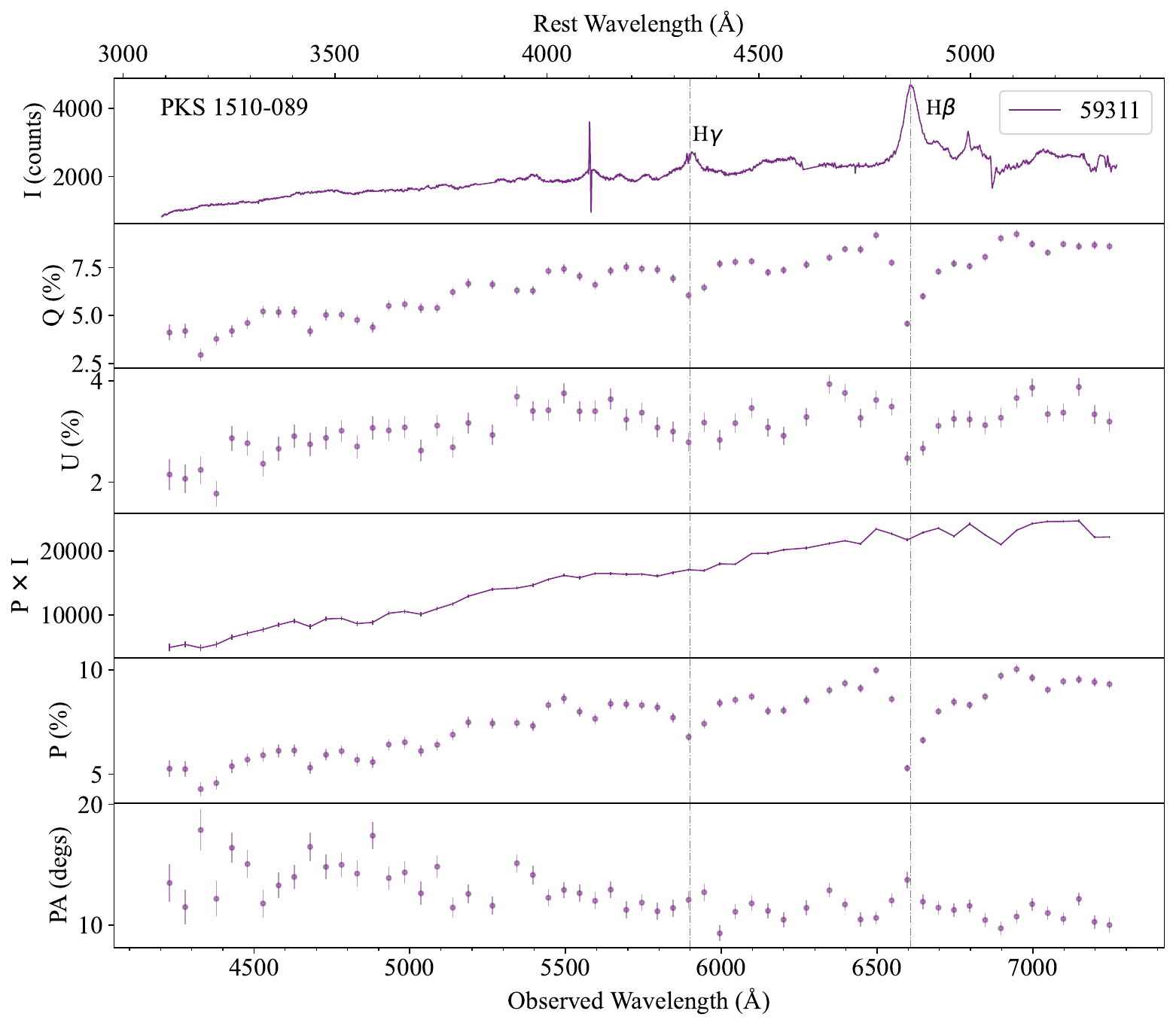}{\columnwidth}{(b)}}
\caption{Example of the total flux spectra, \textit{Q} and \textit{U} normalized Stokes parameters, polarized flux, polarization percentage and PA measurements for an arbitrary observation of PKS 0637--75 (a) and PKS 1510--089   (b). Vertical gray dotted--dashed lines mark Mg II, H$\gamma$, and H$\beta$. }
\label{wapol}
\end{figure}

Continuum polarization measurements were made between 5650 Å and 6260 Å, with the telluric Na I absorption feature around 5900 Å masked, for PKS 0637--75, and between 4480 Å and 5090 Å for PKS 1510--089   due to the lack of any other prominent features within these regions. An example of the total flux spectra, \textit{Q} and \textit{U} normalized Stokes parameters, polarized flux, polarization percentage and PA measurements are shown in Figure~\ref{wapol} for PKS 0637--75 (a) and PKS 1510--089 (b).

After the standard CCD reduction procedures were carried out, we continuum-normalized the spectra by dividing out the modeled continuum from the overall spectra using a polynomial of degree 2, as it provided the best overall fit to our data. In doing so, the chip gaps, emission lines, and any noticeable cosmic rays missed from the reduction pipeline were masked so as to not negatively influence the fit.

In PKS 0637--75, we concentrate on modeling a relatively narrow band between 4500 Å and 5000 Å in the observed frame. The spectrum consists of the continuum, Fe II pseudocontinuum \citep{bruhw,fetemp}, and the Mg II emission line. In PKS 1510--089, we concentrate on modeling the two main broad emission lines, H$\gamma$ and H$\beta$, between 5800 Å and 6800 Å in the observed frame. The Mg II line is treated as a singlet and for all three broad emission lines the kinematic shape is modeled as a single Gaussian using the \texttt{lmfit} package to minimize the $\chi^2$ statistic \citep{lmfit} and obtain the Gaussian parameters plus error used for further analysis.

As a first step to determine if the broad emission lines of our two blazars show variability in polarization levels, we checked if the lines can be seen in polarized light. To do so, we took the nonnormalized spectral intensities, \textit{I}, and multiplied them by the polarization percentage, \textit{P}, to obtain polarized spectra, \textit{P} $\times$ \textit{I}. To increase the signal-to-noise ratio, we bin the full spectra by 50 Å wide bins (Figure~\ref{allspec}, right). After this visual inspection, to further constrain an upper limit of the polarization percentage for the emission lines in this study, \textit{$P_L$}, we create a simple model where the line polarization is some fraction of the continuum polarization, \textit{$P_C$}, it is sitting on (Equation (\ref{pl_equation})):

\begin{equation} \label{pl_equation}
    P_L = P_C \: \frac{A}{R}
\end{equation}
\\
We assume the emission line retains its same shape in polarized light as was displayed in the nonpolarized spectra. Under this assumption the fraction \textit{A/R} compares the amplitude of the Gaussian used to fit where an emission line would be in normalized polarized spectra (\textit{A}) to the amplitude of a Gaussian used to fit the emission line in normalized nonpolarized spectra (\textit{R}), keeping the centroid and $\sigma$ of the Gaussian fixed to what their values were in the non-polarized emission line fit. Standard error propagation was followed to obtain the 1$\sigma$ error values. Negative (nonphysical) polarization values may be returned in this model when a trough is encountered instead of a peak near the emission line position in the polarized spectra.

We measure the equivalent width (EW) of Mg II, H$\gamma$, and H$\beta$ by using the code \texttt{PHEW} (PytHon Equivalent Widths; \citealt{phew}). Within \texttt{PHEW}, we use \texttt{EW.py} to calculate the EW of an emission or absorption line for a given spectrum using \texttt{PySpecKit}, which takes 13 input parameters to find a best-fit EW value and its associated error via Monte Carlo (MC) iterations. Parameters updated for each emission line in the \texttt{equivalent\_width} function include: \texttt{spec}, a list of the spectrum that includes the wavelength, flux, and flux error of the spectrum; \texttt{bandloc}, the value specifying the central location of the spectral line to be measured in Å; \texttt{xmin,xmax}, which specify the wavelength space region of interest around the emission line; \texttt{exclude\_min,exclude\_max} values that specify the start and stop wavelength range of the emission line; \texttt{n}, the number of MC iterations to go through; and \texttt{blorder}, the order of the polynomial used to fit the pseudocontinuum. For all three emission lines studied here, \texttt{n} and \texttt{blorder} used were 500 and 2 respectively. For Mg II we used 4632, 4489, 4840, 4590, and 4720 Å for \texttt{bandloc}, \texttt{xmin,xmax}, and \texttt{exclude\_min,exclude\_max} respectively. We do not use the center value output from the \texttt{lmfit} single-Gaussian fitting for the \texttt{bandloc} values here as they do not create an accurate fit to the emission line when using \texttt{PHEW} to get the EW values; the Gaussian has some additional structure, so the fit is off from the actual centroid location by tens of angstroms in each instance. For H$\gamma$ we used the center value output from the \texttt{lmfit} Gaussian fit for each observation as the \texttt{bandloc} value, with 5800, 6100, 5850, 5970 Å for \texttt{xmin,xmax}, and \texttt{exclude\_min,exclude\_max} respectively. For H$\beta$, we also used the center value output from the \texttt{lmfit} Gaussian fit for each observation as the \texttt{bandloc} value, with 6400, 6800, 6550, 6670 Å for \texttt{xmin,xmax}, and \texttt{exclude\_min,exclude\_max} respectively. Tables \ref{0637ew} and \ref{1510ew} give the calculated EW and errors for Mg II and H$\gamma$ and H$\beta$. Generally, the EW of the Balmer emission lines follow an inverse relationship to the ASAS-SN \textit{g}-flux, showing a decrease as the flux increases.

\begin{deluxetable}{c|cc}
\tablecaption{PKS 0637--75 Mg~II EW Measurements and Associated Error}\label{0637ew}
\tablehead{\colhead{} & \colhead{} & \colhead{Equivalent Width}  \\
     \cline{3-3}
     \colhead{} & \colhead{} & \colhead{(Å)}\\
      \cline{3-3}
\colhead{Date} & \colhead{MJD} & \colhead{Mg~II}}
\startdata
     20190223 & 58537 & 21.4 $\pm$ 0.6 \\
     20190301 & 58543 & 21.4 $\pm$ 0.6 \\
     20190304 & 58546 & 21.5 $\pm$ 0.6 \\
     20191105 & 58792 & 20.3 $\pm$ 0.8 \\
     20191130 & 58817 & 20.2 $\pm$ 0.8 \\
     20191221 & 58838 & 18.8 $\pm$ 0.8 \\
     20191230 & 58847 & 20.2 $\pm$ 1.0 \\
     20200114 & 58862 & 19.0 $\pm$ 1.0 \\
     20200125 & 58873 & 19.4 $\pm$ 0.9 \\
     20201213 & 59196 & 21.7 $\pm$ 0.7 \\
     20210107 & 59221 & 4.7 $\pm$ 0.1  \\
     20210131 & 59245 & 20.9 $\pm$ 0.7 \\
     20210209 & 59254 & 21.6 $\pm$ 0.6 \\
     20210304 & 59277 & 23.1 $\pm$ 0.9  
\enddata
\tablecomments{Absolute value of EW measurements are shown.}
\end{deluxetable}

\begin{deluxetable}{cccc}
\tablecaption{PKS 1510--089   H$\gamma$ and H$\beta$ EW Measurements and associated Error}\label{1510ew}
\tablehead{\colhead{} & \colhead{} & \multicolumn{2}{c}{ Equivalent Width}  \\
     \cline{3-3} \cline{4-4} 
     \colhead{} & \colhead{} & \multicolumn{2}{c}{ (Å)}\\
     \cline{3-3} \cline{4-4} 
\colhead{Date} & \colhead{MJD} & \colhead{H$\gamma$ } & \colhead{H$\beta$}}
\startdata
     20190306 & 58548 & 17.9 $\pm$ 0.6 & ... \\
     20190413 & 58586 & 9.1 $\pm$ 0.3 & ... \\
     20200319 & 58927 & 15.4 $\pm$ 0.5 & ... \\
     20210208 & 59253 & 22.2 $\pm$ 0.8 & 67.2 $\pm$ 0.9 \\
     20210407 & 59311 & 17.0 $\pm$ 0.6 & 47.3 $\pm$ 0.5  \\
     20210416 & 59320 & 22.8 $\pm$ 0.7 & 60.7 $\pm$ 0.7 \\
     20210506 & 59340 & 15.2 $\pm$ 0.6 & 40.8 $\pm$ 0.5 \\
     20210510 & 59344 & 8.3 $\pm$ 0.3 & 23.3 $\pm$ 0.3 \\
     20210512 & 59346 & 11.4 $\pm$ 0.5 & 29.5 $\pm$ 0.4 \\
     20210517 & 59351 & 19.6 $\pm$ 0.6 & 50.7 $\pm$ 0.6 \\
     20210530 & 59364 & 11.4 $\pm$ 0.6 & 29.8 $\pm$ 0.5 \\
     20210606 & 59371 & 8.2 $\pm$ 0.4 & 22.5 $\pm$ 0.4 \\
     20210612 & 59377 & 3.0 $\pm$ 0.7 & 13.9 $\pm$ 0.5 \\
     20210707 & 59402 & 10.2 $\pm$ 0.5 & 23.2 $\pm$ 0.4 \\
     20210711 & 59406 & 5.2 $\pm$ 0.4 & 12.1 $\pm$ 0.3 \\
     20210714 & 59409 & 5.1 $\pm$ 0.3 & 13.4 $\pm$ 0.2 \\
     20210731 & 59426 & 22.0 $\pm$ 0.7 & 55.5 $\pm$ 0.7 \\
     20210804 & 59430 & 18.6 $\pm$ 0.7 & 54.5 $\pm$ 0.6 \\
     20210824 & 59450 & 27.1 $\pm$ 0.8 &  66.8 $\pm$ 0.7 
\enddata
\tablecomments{Absolute value of EW measurements are shown. The ellipses (...) for the first three observations are due to the H$\beta$ emission line being on a CCD chip gap.}
\end{deluxetable} 
\section{Results} \label{res}

\subsection{Level of Continuum Polarization}

Bright and transient sources that have shown flares during the LAT mission and reach the minimum gamma-ray flux threshold of 1$\times$10$^{-6}$ cm$^{-2}$ s$^{-1}$ are added to the \textit{Fermi} monitored source list which provides daily and weekly flux values of such objects of interest. PKS 1510--089 has crossed the abovementioned flux threshold for continual monitoring, representing one of our gamma-ray loud blazars. PKS 0637--75 on the other hand, while a \textit{Fermi} blazar in that it has been gamma-ray detected by \textit{Fermi} and is included in all of the annual \textit{Fermi} catalogues, has not reached the minimum threshold, thus representing one of our gamma-ray quiescent blazars.

For PKS 0637--75, the continuum region was defined between 5650 Å and 6260 Å in the observer's frame to avoid the underlying Fe II at the wavelength of Mg II, as well as any other spectral features. The level of continuum polarization ranges from a minimum of 1.4\% $\pm$ 0.1\% to a maximum of 4.0\% $\pm$ 0.2\%, with an average value of 2.5\% $\pm$ 0.1\%. For PKS 1510--089, the continuum region was defined between 4200 Å and 4900 Å for the first three observations and between 4480 Å and 5090 Å for the remaining observations, as H$\beta$ fell on a chip gap for the first few observing windows. The level of continuum polarization ranges from a minimum of 1.8\% $\pm$ 0.1\% to a maximum of 21.4\% $\pm$ 0.1\%, with an average value of 7.5\% $\pm$ 0.1\%.  

\begin{figure}[ht]
    \centering
    \includegraphics[width = \columnwidth]{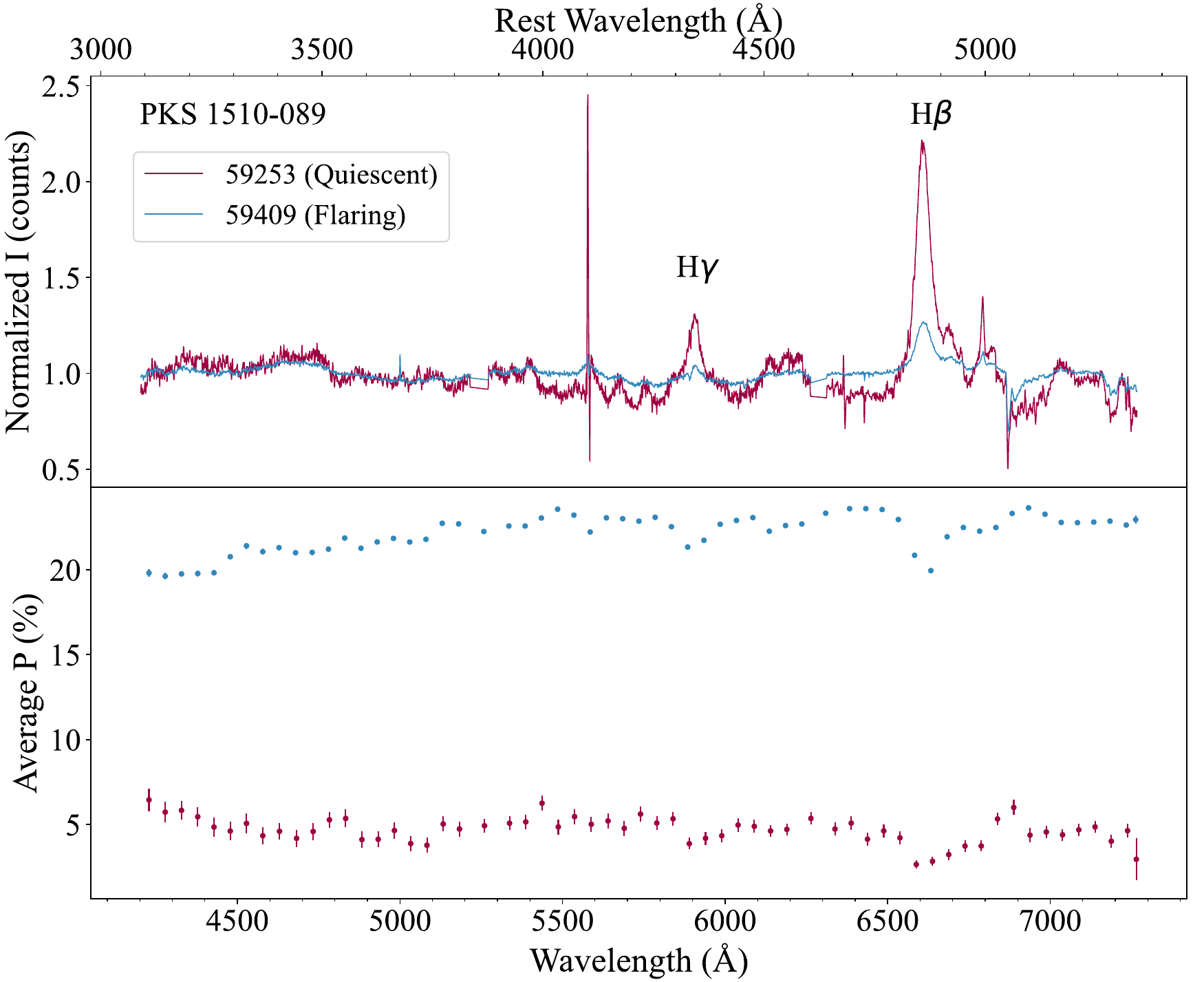}
    \caption{Intensity and polarization variability during optical quiescent (59253) and flaring (59409) periods of PKS 1510--089. We are able to observe the change in thermal to nonthermal emission dominance in the transition from the quiescent to flaring state, with a noticeable decrease in H$\gamma$ and H$\beta$ line strength compared to the continuum and an approximately 4$\times$ increase in overall polarization.}
    \label{flare_quies}
\end{figure}

Continuum polarization for PKS 1510--089 shows stronger variability than for PKS 0637--75. In particular, the highest levels of polarization ($\sim$17\%-21\%) reached by PKS 1510--089 happens between MJDs 59406 and 59409, which corresponds to an optical flare as seen with ASAS-SN (see (a) and (c) of Figure~\ref{1510_ahar_horiz}). We are able to clearly detect a change in the dominant emission processes before (MJD 59253, purple) and near (MJD 59409, blue) the optical flaring period (Figure~\ref{flare_quies}); the emission mechanism changes from thermally dominated, low polarization to nonthermal synchrotron dominance with high levels of polarization detected and smaller EWs of both emission lines.

In both blazars, there is modest evidence for a wavelength dependence of the polarization, with the polarization fraction appearing to rise at redder or bluer wavelengths depending on the date of observation. For those with an increased polarization fraction at redder wavelengths, the continuum is contaminated by bluer unpolarized disk and BLR emission, causing the dilution of the polarization signal at the shorter wavelengths. The dominance of jet or disk emission is reflected in this wavelength dependence, particularly in the quiescent and flaring states of PKS 1510--089 (Figure \ref{flare_quies}). When the synchrotron jet emission is enhanced compared to the thermal AD, polarization levels are high and the wavelength dependence is more noticeable. When the thermal emission is more dominant, as is seen in PKS 0637--75 and the quiescent state of PKS 1510--089, the polarization fraction is low and the wavelength dependence is marginally detected.

\begin{deluxetable}{cccc}
\tablecaption{PKS 0637--75 Continuum and Mg~II Polarization Percentage and Associated 1-$\sigma$ Error} \label{063pol}
\tablehead{\colhead{} & \colhead{} &  \multicolumn{2}{c}{Polarization}\\
     \cline{3-3} \cline{4-4} 
     \colhead{} & \colhead{} & \multicolumn{2}{c}{ (\%)}\\
    \cline{3-3} \cline{4-4} 
     \colhead{Date} & \colhead{MJD} & \colhead{Continuum} & \colhead{Mg~II}} 
\startdata
     20190223 & 58537 & 1.5 $\pm$ 0.1 & 0.3 $\pm$ 0.7 \\
     20190301 & 58543 & 1.5 $\pm$ 0.1 & 0.4 $\pm$ 0.5 \\
     20190304 & 58546 & 1.4 $\pm$ 0.1 & 0.6 $\pm$ 0.4 \\
     20191105 & 58792 & 2.8 $\pm$ 0.1 & 1.4 $\pm$ 1.3 \\
     20191130 & 58817 & 2.8 $\pm$ 0.1 & 0.7 $\pm$ 1.2 \\
     20191221 & 58838 & 2.9 $\pm$ 0.1 & 2.5 $\pm$ 0.9 \\
     20191230 & 58847 & 3.1 $\pm$ 0.1 & \textless \: 1.0$^*$ \\
     20200114 & 58862 & 4.0 $\pm$ 0.2 & 2.1 $\pm$ 1.0 \\
     20200125 & 58873 & 3.2 $\pm$ 0.1 & 1.3 $\pm$ 1.3 \\
     20201213 & 59196 & 2.3 $\pm$ 0.1 & 2.3 $\pm$ 0.9 \\
     20210107 & 59221 & 2.3 $\pm$ 0.1 & 0.6 $\pm$ 0.7 \\
     20210131 & 59245 & 2.3 $\pm$ 0.1 & 1.4 $\pm$ 1.2 \\
     20210209 & 59254 & 2.2 $\pm$ 0.1 & \textless \: 0.2$^*$ \\
     20210304 & 59277 & 2.9 $\pm$ 0.1  & \textless \: 0.5$^*$ 
\enddata
\tablecomments{Polarization values with an * denote upper limit values.}
\end{deluxetable}

\begin{deluxetable}{ccccc}
\tablecaption{PKS 1510--089 Continuum and Emission Line Polarization Percentage and 1-$\sigma$ Error} \label{1510pols}
\tablehead{\colhead{} & \colhead{} &  \multicolumn{3}{c}{Polarization}\\
     \cline{3-3} \cline{4-4} \cline{5-5}
     \colhead{} & \colhead{} & \multicolumn{3}{c}{ (\%)}\\
    \cline{3-3} \cline{4-4} \cline{5-5}
     \colhead{Date} & \colhead{MJD} & \colhead{Continuum} & \colhead{H$\gamma$} & \colhead{ H$\beta$}} 
\startdata
     20190306 & 58548 & 2.1 $\pm$ 0.1 & 0.1 $\pm$ 1.0 & \textless \: 0.1$^*$ \\
     20190413 & 58586 & 4.5 $\pm$ 0.1 & \textless \: 1.2$^*$ & 0.4 $\pm$ 0.5 \\
     20200319 & 58927 & 7.3 $\pm$ 0.1 & \textless \: 0.7$^*$ & 0.7 $\pm$ 0.5\\
     20210208 & 59253 & 4.5 $\pm$ 0.1 & \textless \: 1.4$^*$ & 0.3 $\pm$ 0.9 \\
     20210407 & 59311 & 6.0 $\pm$ 0.1 & \textless \: 0.4$^*$ & \textless \: 0.1$^*$\\
     20210416 & 59320 & 3.0 $\pm$ 0.1 & 0.2 $\pm$ 1.8 & 0.9 $\pm$ 0.6 \\
     20210506 & 59340 & 5.0 $\pm$ 0.1 & \textless \: 0.2$^*$ & 0.6 $\pm$ 0.6 \\
     20210510 & 59344 & 13.2 $\pm$ 0.1 &  0$^{**}$ & 0.9 $\pm$ 0.8\\
     20210512 & 59346 & 14.8 $\pm$ 0.1 & 0$^{**}$ & 0.8 $\pm$ 0.8 \\
     20210517 & 59351 & 3.2 $\pm$ 0.1 & 2.1 $\pm$ 1.3 & 0.4 $\pm$ 0.4 \\
     20210530 & 59364 & 6.5 $\pm$ 0.1 & 2.2 $\pm$ 1.3 & 0.1 $\pm$ 0.9 \\
     20210606 & 59371 & 1.8 $\pm$ 0.1 & 0.3 $\pm$ 1.9 & 1.4 $\pm$ 1.0 \\
     20210612 & 59377 & 12.2 $\pm$ 0.1 &  0$^{**}$ & 3.1 $\pm$ 4.0 \\
     20210707 & 59402 & 9.1 $\pm$ 0.1 & 0.3 $\pm$ 1.1 & 0.2 $\pm$ 0.7 \\
     20210711 & 59406 & 17.0 $\pm$ 0.1 &  0$^{**}$ & \textless \: 0.5$^*$ \\
     20210714 & 59409 & 21.4 $\pm$ 0.1 &  0$^{**}$ & 1.8 $\pm$ 1.5 \\
     20210731 & 59426 & 3.2 $\pm$ 0.1 & 4.3 $\pm$ 1.1 & 0.6 $\pm$ 0.5 \\
     20210804 & 59430 & 3.2 $\pm$ 0.1 & 5.8 $\pm$ 6.8 & \textless \: 0.7$^*$ \\
     20210824 & 59450 & 3.4 $\pm$ 0.1 & 0.2 $\pm$ 1.3 & 0.5 $\pm$ 0.7 
\enddata
\tablecomments{Polarization values with an * denote upper limit values, and those with ** denote the $1\sigma$ upper limit was a (nonphysical) negative value, so we mark the line as not polarized. The measured values with uncertainties are shown in Figure \ref{line_con_wapol}.}
\end{deluxetable}

\subsection{Level of Emission Line Polarization }

\begin{figure}[ht]
\gridline{\fig{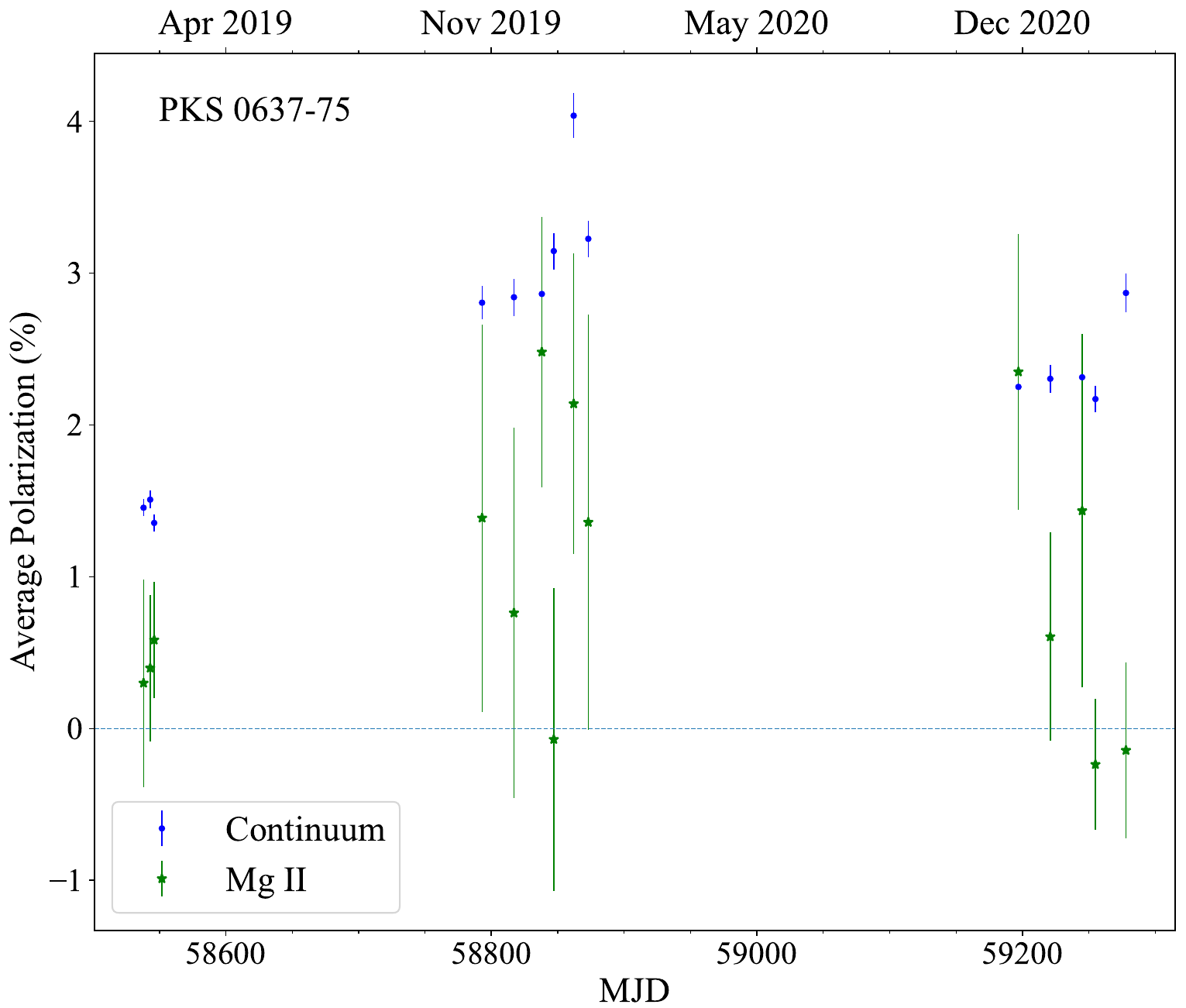}{\columnwidth}{(a)}}
\gridline{\fig{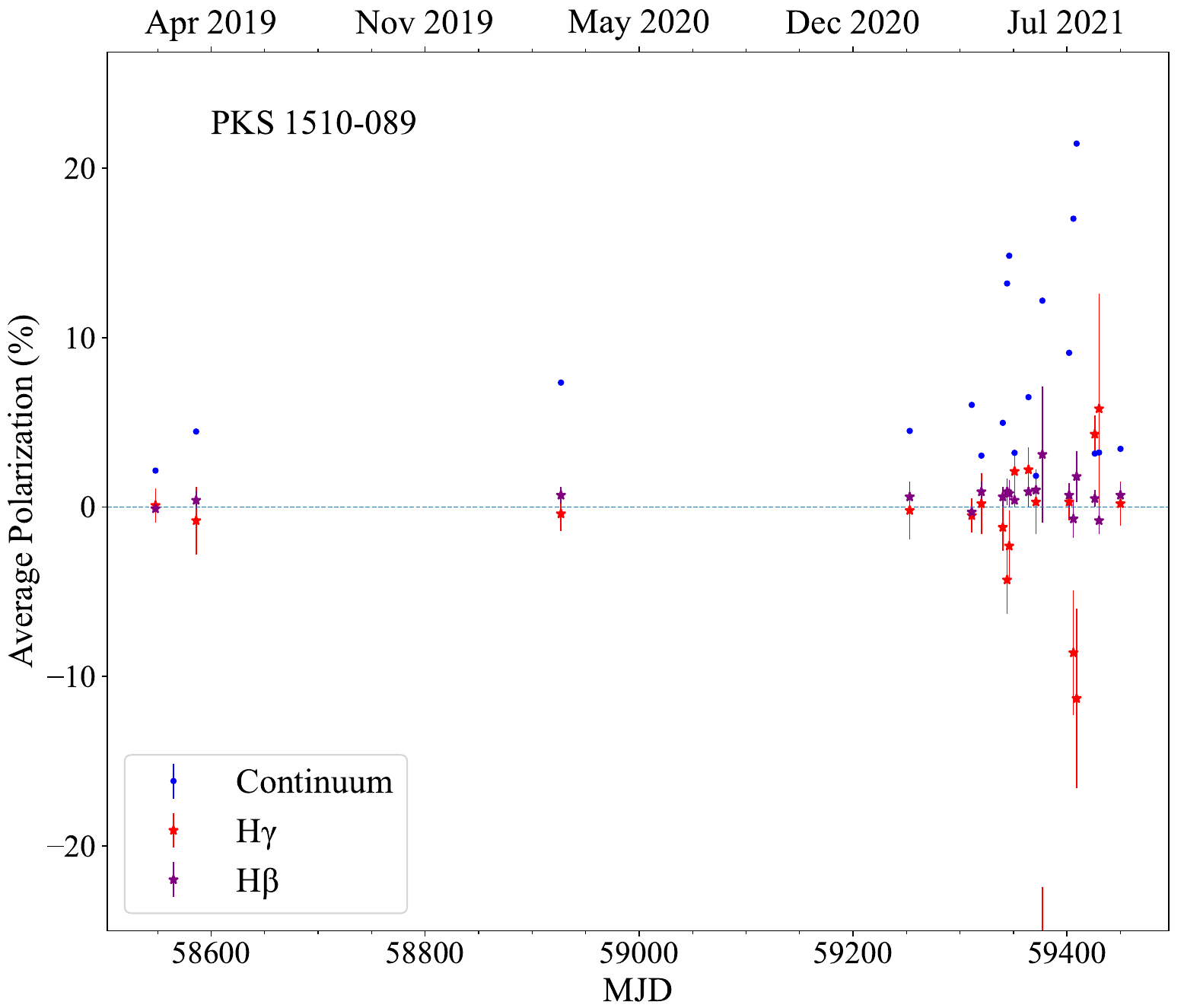}{\columnwidth}{(b)}}
\caption{(a) PKS 0637--75 continuum and Mg~II average polarization percentage as a function of time. Variability in continuum polarization is detected across the full observation period, with slight polarization of the emission line potentially detected. (b) PKS 1510--089 continuum, H$\gamma$ and H$\beta$ average polarization percentage as a function of time. This shows a more polarized continuum as compared to PKS 0637--75, and while the continuum polarization increases and varies over the full observation period, the emission line polarization values are nonpolarized in all observations.}
\label{line_con_wapol}
\end{figure}

From our initial visual inspection of \textit{P}$\times$\textit{I} in Figure~\ref{allspec} (right), neither PKS 0637--75 nor PKS 1510--089 display noticeable features at the wavelengths of the emission lines Mg II, H$\gamma$, or H$\beta$. In Figure~\ref{wapol_all}, there is an observable depolarization dip in the average polarization levels most notably at the wavelengths corresponding to the centroid of the H$\gamma$ and H$\beta$ lines, marked with the vertical dotted-dashed gray lines. Moving to our simple model used to find \textit{$P_L$} for each emission line, we find the following polarization level limits: For Mg II, the minimum and maximum \textit{$P_L$} values are \textless 0.2\% and 2.5\% $\pm$ 0.9\%. H$\gamma$ is consistent with zero polarization throughout, with a maximum \textit{$P_L$} value of 5.8\% $\pm$ 6.8\%. Similarly, the minimum and maximum \textit{$P_L$} values for H$\beta$ are \textless 0.5\% and 3.1\% $\pm$ 4.0\%. Mg~II shows a weak level of polarization while the calculated line polarization percentages are almost all identically similar to zero within their 1$\sigma$ error bars for the Balmer lines, i.e. these broad emission lines are not polarized. Tables \ref{063pol} and \ref{1510pols} give the calculated continuum and emission line polarization for PKS 0637--75 and PKS 1510--089 respectively; see Figure~\ref{line_con_wapol} for visual representation.

\subsection{Polarization Angle}

Table \ref{pa} and Figure~\ref{wapa} show the variation with time in the PA for both blazars. PKS 1510--089 shows a range of values between 0$^{\circ}$ and 175$^{\circ}$ and PKS 0637--75 has a few distinct clumps of values at around 0$^{\circ}$--30$^{\circ}$, 80$^{\circ}$  and 150$^{\circ}$--200$^{\circ}$. The continuum polarization angle of PKS 0637--75 ranges from -3$^{\circ}$.3 $\pm$ 1$^{\circ}$.2 to 190$^{\circ}$.4 $\pm$ 1$^{\circ}$.0, with an average value of 127$^{\circ}$.8 $\pm$ 1$^{\circ}$.0. For PKS 1510--089, the continuum polarization angle ranges from -1$^{\circ}$.2 $\pm$ 0$^{\circ}$.7 to 180$^{\circ}$.5 $\pm$ 0$^{\circ}$.8, with an average value of 95$^{\circ}$.9 $\pm$ 0$^{\circ}$.5. 

Near the time of enhanced optical flux between MJDs 59375 and 59415 for PKS 1510--089, the total average polarization percentage shows elevated levels and the PA changes by about 120$^{\circ}$. During this time, the H$\beta$ and H$\gamma$ line polarization is nominally 0\% within their errors.

\begin{deluxetable}{ccc}
    \tablecaption{Average Continuum Polarization Angle and Associated 1-$\sigma$ Error for Both PKS 0637--75 and PKS 1510--089}\label{pa}
    \tablehead{\colhead{} & \colhead{} & \colhead{Polarization Angle} \\
    \colhead{Date} & \colhead{MJD} & \colhead{(deg)}}
    \startdata
    \multicolumn{3}{c}{\emph{PKS 0637--75}} \\
    \cline{1-3} 
     20190223 & 58537 & -3.3 $\pm$ 1.2  \\
     20190301 & 58543 & 16.9 $\pm$ 0.9 \\
     20190304 & 58546 & 168.0 $\pm$ 1.1  \\
     20191105 & 58792 & 170.8 $\pm$ 1.0 \\
     20191130 & 58817 & 164.8 $\pm$ 0.9   \\
     20191221 & 58838 & 174.4 $\pm$ 0.9   \\
     20191230 & 58847 & 167.1 $\pm$ 1.0 \\
     20200114 & 58862 & 66.1 $\pm$ 1.0 \\
     20200125 & 58873 & 190.4 $\pm$ 1.0 \\
     20201213 & 59196 & 174.2 $\pm$ 1.0  \\
     20210107 & 59221 & 7.0 $\pm$ 1.1 \\
     20210131 & 59245 & 158.7 $\pm$ 1.0\\
     20210209 & 59254 & 157.4 $\pm$ 1.0  \\
     20210304 & 59277 & 177.2 $\pm$ 1.0 \\
    \multicolumn{3}{c}{\emph{PKS 1510--089}} \\
    \cline{1-3} 
     20190306 & 58548 & 180.5 $\pm$ 0.8 \\
     20190413 & 58586 & 91.5 $\pm$ 0.2 \\
     20200319 & 58927 & 120.4 $\pm$ 0.3 \\
     20210208 & 59253 & -1.2 $\pm$ 0.7 \\
     20210407 & 59311 & 14.4 $\pm$ 0.3 \\
     20210416 & 59320 & 70.4 $\pm$ 0.9 \\
     20210506 & 59340 & 135.4 $\pm$ 0.5 \\
     20210510 & 59344 & 164.6 $\pm$ 0.1 \\
     20210512 & 59346 & 22.0 $\pm$ 0.1 \\
     20210517 & 59351 & 61.3 $\pm$ 0.7 \\
     20210530 & 59364 & 98.5 $\pm$ 0.4\\
     20210606 & 59371 & 44.3 $\pm$ 0.9 \\
     20210612 & 59377 & 63.3 $\pm$ 0.3 \\
     20210707 & 59402 & 43.1 $\pm$ 0.2 \\
     20210711 & 59406 & 171.3 $\pm$ 0.1 \\
     20210714 & 59409 & 166.3 $\pm$ 0.1 \\
     20210731 & 59426 & 125.1 $\pm$ 0.7 \\
     20210804 & 59430 & 142.3 $\pm$ 0.8 \\
     20210824 & 59450 & 51.1 $\pm$ 0.7 
\enddata
\end{deluxetable}

\subsection{Broadband Spectral Energy Distribution Modeling}

PKS 0637--75, unlike PKS 1510--089, is a source that does not exhibit much variability generally. To better understand what is contributing to the observed flux and polarization of PKS 0637--75, we compiled the SED
for semicontemporaneous observations between ASAS-SN \textit{g}-band and Swify-UVOT photometry corresponding to our SALT 20190303 epoch of observation, with NED (NASA/IPAC Extragalactic Database (NED) \citep{ned} archival data from 2005 onward. A similar analysis of PKS 1510--089 has been done by \cite{aharonian+23}, so we refer to that work for a detailed study of its SED.
 
The leptonic single-zone blazar model of \cite{bot13} has been used to produce a fit by eye to the broadband SED of PKS 0637--75. A summary of this model is given here; see \cite{bot13} for a more detailed and thorough description. This model follows a homogeneous one-zone framework where a power-law distribution of ultrarelativistic electrons is injected into a spherical emission region of radius $R$. The emission region moves with a constant speed $\beta_{\Gamma}$\textit{c} along the jet, which corresponds to the bulk Lorentz factor $\Gamma$.

The cooling of the electron distribution is influenced by synchrotron and Compton emission processes. Synchrotron emission is determined by a tangled magnetic field of strength \textit{B}. The model accounts for Compton scattering involving the synchrotron radiation field (synchrotron self Compton, SSC) and external radiation fields (external Compton, EC), including direct AD emission --- EC (disk) --- and AD emission reprocessed by the BLR, represented numerically by an isotropic external radiation field with a blackbody spectrum of temperature $k T_{\rm BLR} = 10^4$~K --- EC (BLR).

The code self-consistently computes an equilibrium electron distribution considering particle injection/acceleration, escape on an energy-independent escape time scale $t_{\rm esc} = \eta_{\rm esc} \, R/c$ with $\eta_{\rm esc} \ge 1$, and radiative cooling; evaluates the kinetic jet power $L_e$ corresponding to the final electron population in the emission region and the magnetic field (Poynting flux) power $L_B$; and calculates the energy partition ratio $\epsilon_{Be}$ $\equiv$ $L_B/L_e$.

Observational parameters used to constrain the model include the redshift (\textit{z} = 0.653), the black hole mass \citep[M$_{BH}$ $\sim$ 10$^9$ M$_{\odot}$;][]{xiong14, ito21}, the apparent superluminal speed \citep[$\beta_{\perp,app}$ $\sim$ 13;][]{godfrey12}, and the AD and BLR luminosities \citep[$L_{AD}$ $\sim$ 10$^{47}$ erg s$^{-1}$, $L_{BLR}$ $\sim$ 10$^{45}$ erg s$^{-1}$;][]{xiong14}. The fit parameters adjusted during the fit-by-eye procedure include the low-energy and high-energy cutoffs of the injected electron spectrum ($\gamma_{min}$, $\gamma_{max}$), the electron injection spectral index (q$_e$), the emission region radius ($R$), the magnetic field strength (\textit{B}), the distance of the emission region from the black hole (\textit{z$_0$}), the bulk Lorentz factor ($\Gamma$), the observing angle ($\theta_{obs}$) in the observer's frame, the BLR radiation field black-body temperature (T$_{BB}$), and the BLR radiation field energy density (u$_{BB}$). The minimum variability time scale corresponding to the light-crossing time scale (t$_{var,min}$), the kinetic power in relativistic electrons in the AGN frame ($L_e$), the power carried in the magnetic field ($L_B$), and the energy partition parameter ($\epsilon_{Be}$) are computed from the other model parameters and the resulting equilibrium electron distribution. 

\begin{figure*}[ht]
    \centering
    \includegraphics[width = 0.9\textwidth]{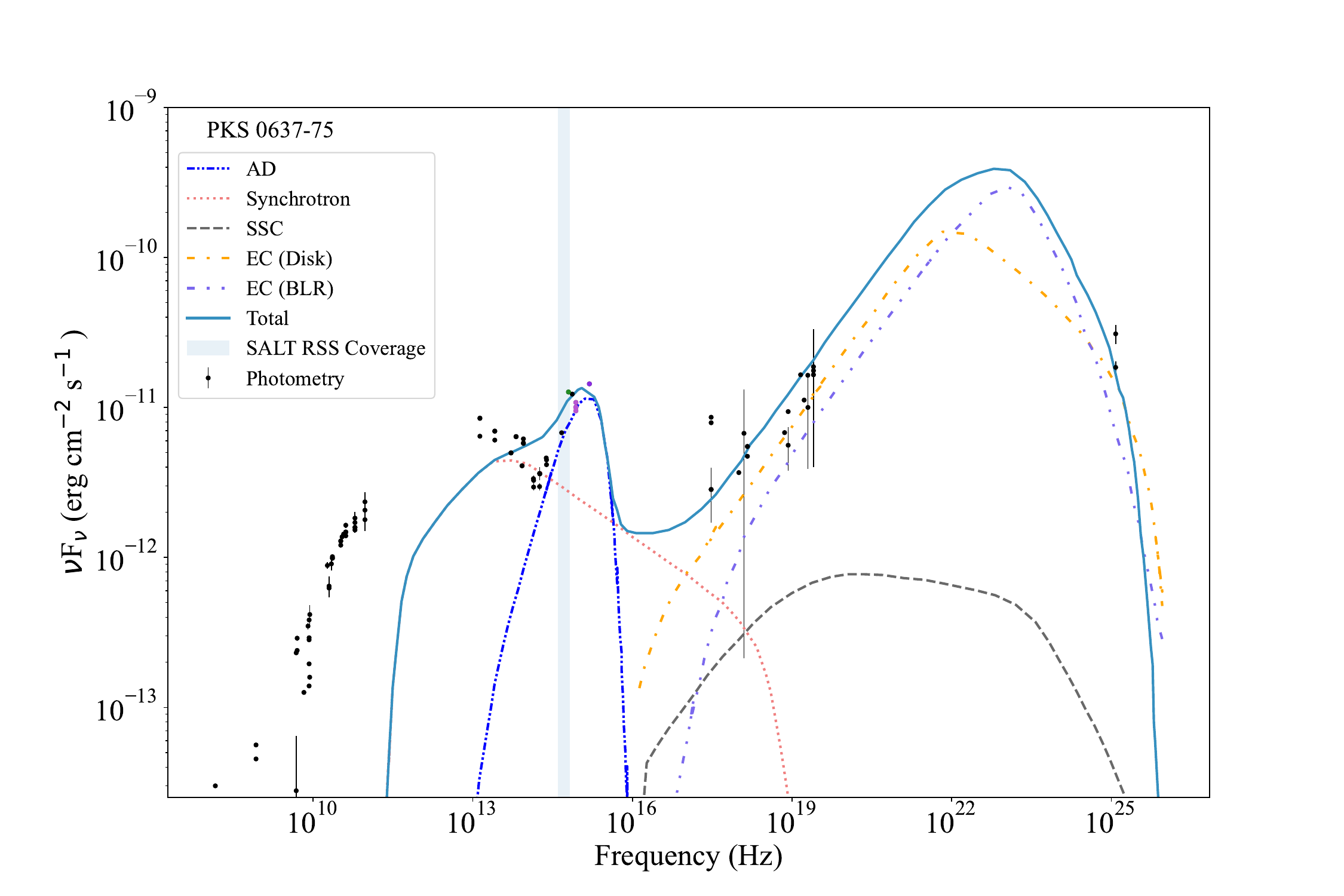}
    \caption{Single-zone leptonic model fit to the broadband SED of PKS 0637--75 using the code of \cite{bot13}. Photometry data (points with error bars) are as follows: archival multiwavelength data points from NED (black), quasi-simultaneous ASAS-SN \textit{g}-band (green) and Swift-UVOT \textit{u} (orchid) and \textit{uvw2}bands (blue violet). The plotted radiation components include: ADflux (blue double dotted--dashed), synchrotron emission (coral dotted), synchrotron self-Compton emission (gray dashed), external Compton emission from the AD (EC (Disk); orange dotted--dashed) and external Compton emission from the broad-line region (EC (BLR); light purple loosely double dotted--dashed), and the best fit of the total components (teal solid). The light shaded box around 10$^{15}$ Hz indicates the frequency coverage of our spectropolarimetry data from the RSS on SALT. While the fit-by-eye is rough and nonunique, it is apparent that we are detecting mainly thermal emission in our SALT observations.}
    \label{fig:0637_MB13sed}
\end{figure*}

A representative broadband SED fit is plotted in Figure~\ref{fig:0637_MB13sed} with the leptonic model fit parameters listed in Table \ref{sed_params}. However, due to the substantial number of free parameters and the fact that some of the SED photometric measurements were taken at different times, the fit is rough and nonunique. In order to reduce the degeneracies in the fit procedure, we aimed for a fit with exact equipartition between relativistic electrons and magnetic fields, which we were able to achieve. On the basis of the SED fit, we determined that the optical emission is dominated by the thermal disk, though the listed parameters are only indicative and no strict conclusions on the physical conditions in the emission region should be drawn from it.

\begin{deluxetable}{cc}[ht]
\tablecaption{Model Parameters for the SED Fit of PKS 0637--75} \label{sed_params}
\tablehead{\colhead{Parameter} & \colhead{Value}}
\startdata
     M$_{BH}$ (M$_{\odot}$) & 2.5 $\times$ 10$^9$ \\
     $L_{disk}$ (erg s$^{-1}$) & 8 $\times$ 10$^{46}$  \\
     $L_{BLR}$ (erg s$^{-1}$) & $\sim$ 10$^{45}$  \\
     $\gamma_{min}$ & 700\\
     $\gamma_{max}$ & 2 $\times$ 10$^5$\\
     \textit{q$_e$} & 2.5 \\
     \textit{R} (cm) & 1.5 $\times$ 10$^{16}$ \\
     \textit{B} (G)  & 2  \\
     z$_0$ (pc) & 0.06 \\
     $\Gamma$ & 15\\
     $\theta_{obs}$ (deg) & 3.82\\
     T$_{BB}$ (K) & 1 $\times$ 10$^4$  \\
     u$_{BB}$ (erg cm$^{-3}$) & 2 $\times$ 10$^{-2}$ \\
     $L_e$ (erg s$^{-1}$) & 7.60 $\times$ 10$^{44}$  \\
     $L_B$ (erg s$^{-1}$) & 7.59 $\times$ 10$^{44}$  \\
     $\epsilon_{Be}$ & 1.0\\
     t$_{var,min}$ (hr) & 15.3 
\enddata
\end{deluxetable}
\section{Discussion} \label{dis}

The optical emission from blazars is composed of a variety of components including the accretion disk, the broad-line region, and synchrotron emission from the jet. Each of the emission mechanisms associated with these regions contributes differently to the polarization characteristics of the emitted radiation \citep{kovalev20}. The mechanisms inducing polarization in AGNs can be divided into internal (central parsec and smaller scales) and external (greater-than-parsec scales). Polarization due to the influence of magnetic fields close to the SMBH, radiation transfer from the AD, the synchrotron jet radiation, and electron scattering in the hot corona all contribute to the internal polarization mechanisms. Equatorial scattering at the torus and polar scattering at the ionization cone make up the external polarization mechanisms \citep{Shablov19}.

Polarized emission lines are typically due to scattering by material through the above mentioned external mechanisms, where the number and location of the scattering regions determine the observed polarization properties. If equatorial scattering is the dominant mechanism, a couple of characteristic polarization signatures would be observed -- mainly a dip in the polarization degree and an S-shaped PA swing along the emission line profiles \citep{Shablov23}. Of the two, we do see a dip in the percentage of polarization at each emission line core region, but there is no evidence for any swing of the PA across the line (Figure~\ref{wapa}).

The linear polarization that is associated with optically thin synchrotron radiation depends on the structure of the magnetic fields in the emitting region and on the emitting electrons' energy distribution \citep{itoh16}. Additionally, polarization from nonthermal electrons in an anisotropic magnetic field should vary with the total flux of an object as magnetic field configurations evolve, giving rise to synchrotron polarization that is strongly variable and different from the polarization of emission lines \citep{Marin19}. Especially evident for PKS 1510--089 , the polarization fraction spanning the 2--3 yr of our study displays variability with flux and is different from that of the emission lines, which are consistently unpolarized. 

The orientation of an AGN with respect to the observer's line of sight can have a strong influence on the polarization detection as well. Single SMBHs surrounded by coplanar, axisymmetric, or spherically-shaped scattering regions produce low amounts of polarization when viewed close to face-on inclinations \citep{savic+19}. For objects with inclination angles close to 0$^{\circ}$ (i.e. blazars), we have a mostly direct, unobstructed view of the AD, BLR, and relativistic jet. As such, the jet will continue to have a polarization signal due to the intrinsically polarized synchrotron radiation, while the polarization vectors of the disk undergo geometric cancellations, resulting in no net disk polarization in total light \citep{OudHar08} nor in the broad lines.

Various spectropolarimetry studies of quasars have been undertaken with similar results to what we have found for PKS 0637--75  and PKS 1510--089. \cite{kishimoto03} found an absence of broad lines in polarized light for two quasars in their study, suggesting the continuum scattering region potentially is located interior to or cospatially with the BLR \citep[from][]{smith2005}. Additionally, \cite{kishimoto04} found that polarization was confined to the continua for five quasars in their study with depolarized BLR emission and wavelength-independent position angles, suggesting a single source of the observed polarization. For radio-loud quasars studied by \cite{popovic+21}, polarization was not detected in or across the broad emission lines which the authors discuss could have been due to a lack of equatorial scattering, a region of depolarization above the BLR, or an inner equatorial scattering region comparable in size to the BLR. As was discussed for higher inclination quasars in \cite{capetti21}, if a single scattering medium physically close to the BLR is present, geometric cancellation of polarization vectors is possible.

Lower levels of synchrotron polarization during the nonflaring epochs of PKS 1510--089 indicate a less ordered magnetic field during such quiescent states, suggesting more tangled magnetic field lines with different field-line directions that can cancel out \citep{schutte}. The generally low degree of synchrotron polarization in PKS 0637--75  is likely due to the dominant AD emission.

As we see in Figure~\ref{fig:0637_MB13sed}, the upturn at the optical-UV regime in the SED of PKS 0637--75  is well modeled by a Shakura-Sunyaev AD \citep{shak_suny73}. The SED is modelled with an emission region located 0.06 pc down the jet, i.e., within the BLR, hence the dominant external Compton contributions from the disk and BLR (EC (disk), EC (BLR)). The lightly shaded box around 10$^{15}$ Hz illustrates the frequency range covered by the RSS on SALT used for our spectropolarimetry data collection. We see that in our model, SSC is sub-dominant and the optical emission is dominated by thermal processes (direct AD emission) over the nonthermal synchrotron jet. Thus the low degree of polarization of the optical-UV emission in PKS 0637--75 is consistent with dilution by the AD and BLR which are expected to be intrinsically nonpolarized.

From the beginning of 2021 April to mid-June, there exists an overlap of SALT spectropolarimetry observations of PKS 1510--089 between this observing campaign and that of \cite{aharonian+23}, which used H.E.S.S., ATOM, and SALT observations to better understand the primary emission region of PKS 1510--089 (see panels (c) and (d) of Figure \ref{1510_ahar_horiz}). In this study, we observed a drop in the optical continuum polarization and an increase in the EW of broad emission lines H$\gamma$ and H$\beta$ during the quiescent stages of PKS 1510--089, most notably after the optical flaring event around 2021 14 July. After this observation in 2021, our SALT data display significantly lower levels of polarization (less than 4\%). In \cite{aharonian+23}, the polarization level reached a maximum of 12.5\% and minimum of 2.2\% in 2021, similar to the values we obtained from our SALT observations at similar times (within $\pm$ 1--2 days). 

Using the SED + spectropolarimetry model of \cite{schutte}, \cite{aharonian+23} were able to show that contributions from the synchrotron jet, AD, and BLR can explain the observed emission and polarization levels in 2021, whereas for 2022, the drop in polarization level to being consistently below 2\% is consistent with no polarization in the blazar, such that the AD and BLR flux is sufficient to fully explain the optical spectrum. The decrease in polarization levels after 2021 mid-July we observe gives support to the above suggestion that the optical-UV spectrum became dominated by the thermal AD and BLR, while low levels of polarization (4\%) between 2021 mid-July and August may be consistent with only interstellar polarization and no intrinsic polarization in the source, in agreement with what was found by \cite{aharonian+23} in their later observations; see the bottom panels (c) and (d) in Figure \ref{1510_ahar_horiz} where we see in more detail the decline in polarization levels of PKS--1510-089.
\section{Conclusions} \label{con}

We obtained spectropolarimetric observations of FSRQs PKS 0637--75 and PKS 1510--089 using the Southern African Large Telescope during  the 2019--2021 period. Blazar optical emission is composed of the thermal accretion disk and broad-line region and nonthermal synchrotron jet. The connection between these thermal and nonthermal components are explored through the polarization characteristics of the emitted radiation.

Variability in continuum polarization is on the order of approximately half to a few percent on various time scales for PKS 0637--75 and approximately a few to tens of percent on day--week timescales for PKS 1510--089. While we detect variability in the continuum polarization levels, the same cannot be said for the broad emission lines these blazars exhibit. The broad H$\gamma$ and H$\beta$ lines of PKS 1510--089 are not detected in polarized emission and within their errors are consistent with zero polarization. In PKS 0637--75, the low-ionization Mg~II line is not detected in polarized light as well, though it does occasionally demonstrate weak levels of polarization. 

It has been shown for 4C+01.02 that the AD emission can dilute the synchrotron emission toward higher optical frequencies, causing a decrease in the total degree of polarization, as well as a detected decrease in polarization at the frequency of unpolarized emission lines \citep{schutte}. Both of these phenomena are observed in this work -- during the nonflaring period of time when the thermal emission components are dominant, PKS 1510--089 was in a lower polarization state compared to the epoch of nonthermal dominance during the flare. Likewise, there is a noticeable drop in polarization at the wavelengths of the unpolarized emission lines, especially for the Balmer lines.

We conclude that the broad emission lines of PKS 0637--75 and PKS 1510--089 are intrinsically nonpolarized, though geometric cancellations due to the pole-on orientation  potentially exist as well \citep{OudHar08, capetti21} and are causing our nondetection of polarized emission from the lines analyzed here. Additionally, changes in the dominant emission process can lead to continuum polarization variability. The gamma-ray quiet FSRQ PKS 0637--75 is not as variable a source as compared to the gamma-ray loud FSRQ PKS 1510--089 and seems to be consistently dominated by thermal emission in the optical-UV regime, resulting in very low levels of observed polarization. Emission from PKS 1510--089 prior to 2021 mid July was consistent with being associated with a nonthermal synchrotron jet, thermal AD and BLR which then underwent a change to being dominated by the thermal components, as evidenced by the drastic decrease in polarization levels observed in our SALT spectropolarimetric observations and supports the photometric and spectropolarimetric observations from \cite{aharonian+23}. 

\acknowledgements
\noindent

We thank the anonymous reviewer for comments that improved this work. S.A.P. acknowledges support from the Dartmouth Fellowship and Sigma Xi grant G201903158443203 and would like to thank Keighley E. Rockcliffe, Aylin García Soto, John R. Thorstensen, Elisabeth R. Newton, Rujuta A. Purohit, and Emily M. Boudreaux for various mentoring, advice, and conversations that improved the manuscript. R.C.H. acknowledges support from NASA through Astrophysics Data Analysis grant No. 80NSSC23K0485. 

All of the spectropolarimetric observations reported in this paper were obtained with the Southern African Large Telescope (SALT), under proposals 2018-2-SCI-039 (PI: J. Isler), 2019-2-SCI-040 (PI: J. Isler), 2020-2-SCI-017 (PI: S. Podjed), and 2021-1-SCI-027 (PI: S. Podjed) with SALT astronomers Danièl Groenewald \& Lee Townsend--special thanks to the SALT observation team for their diligent communications, data collection, and initial reduction of data for our program study. 

This research has made use of the NASA/IPAC Extragalactic Database (NED) which is operated by the Jet Propulsion Laboratory, California Institute of Technology, under contract with the National Aeronautics and Space Administration. This work includes data collected by the ASAS-SN mission and \textit{Fermi}-LAT mission.  

\facility{SALT (RSS)}
\software{polsalt \citep{polsalt}, lmfit \citep{lmfit}, Astropy \citep{astropy13, astropy18, astropy22}, \texttt{IRAF} \citep{iraf}, PHEW \citep{phew}}

\bibliographystyle{aasjournal}
\bibliography{specpol_refs.bib}

\end{document}